\documentclass{jfm}
\usepackage{graphicx}
\usepackage{epstopdf, epsfig}
\usepackage{amsmath}
\usepackage{cases}
\usepackage{color}
\usepackage{hyperref}
\usepackage{amssymb}
\usepackage{graphics}
\usepackage{subfigure}
\usepackage{textcomp}
\usepackage{bm}
\usepackage{url}
\usepackage{tikz}
\allowdisplaybreaks[1]

\makeatletter
  \newcommand\figcaption{\def\@captype{figure}\caption}
  \newcommand\tabcaption{\def\@captype{table}\caption}
\makeatother

\shorttitle{The HAM for the steep gravity wave}
\shortauthor{X. X. Zhong and S. J. Liao}

\title{On the Limiting Stokes' Wave of Extreme Height in Arbitrary Water Depth}

\author{Xiaoxu Zhong\aff{3}
  \and Shijun Liao\aff{1,2,3}
  \corresp{\email{sjliao@sjtu.edu.cn}},
  }

\affiliation{\aff{1} State Key Laboratory of Ocean Engineering, Shanghai 200240, China
\aff{2} Collaborative Innovative Center for Advanced Ship and Deep-Sea Exploration,\\ Shanghai 200240, China
\aff{3} School of Naval Architecture, Ocean and Civil Engineering\\
Shanghai Jiao Tong University, Shanghai 200240, China
}

\begin{document}

\maketitle

\begin{abstract}

As mentioned by Schwartz (1974) and Cokelet (1977),  it was failed to gain convergent results of limiting Stokes' waves in extremely shallow water by means of  perturbation methods even with the aid of extrapolation techniques such as  Pad\'{e} approximant.  Especially, it is extremely difficult for traditional analytic/numerical approaches to present the wave profile of limiting waves with a sharp crest of $120^\circ$ included angle first mentioned by Stokes in 1880s.   Thus,  traditionally,   different wave models  are used for  waves  in different water depths.   In this paper, by means of the homotopy analysis method (HAM), an analytic approximation method for highly nonlinear equations, we successfully gain convergent results (and  especially the wave profiles) of the limiting Stokes' waves  with this kind of sharp crest in arbitrary water depth, even including solitary waves of extreme form in extremely shallow water, without using any extrapolation techniques.   Therefore,  in the frame of the HAM, the Stokes' wave can be used as a unified theory for all  kinds  of  waves, including periodic waves in deep and intermediate depth, cnoidal waves in shallow water and solitary waves in extremely shallow water.
\end{abstract}


\section{Introduction}

The two-dimensional steady progressive gravity wave is one of the most classic problems in fluid mechanics, which can be tracked back to \citet{StokesOn, Stokes1880Supplement} and was widely studied by lots of researchers \citep{Michell1893, Nekrasov1920, Yamada1957, Yamada1968On, Schwartz1972, Byatt1976On, J1979Numerical, Chen1980Numerical, Olfe1980Some, And1982Strongly, Sulem1983Tracing, Hunter1983Accurate, Vanden1986Steep,  Klopman1990A, Fenton1990, Karabut1998An, Dallaston2010Accurate,  Lushnikov2016Structure, Lushnikov2017New}.  Among analytic approaches for this problem, perturbation methods are used most frequently.  \citet{StokesOn, Stokes1880Supplement} proposed  a  perturbation approach using the first Fourier coefficient, $a_1$, as the perturbation quantity, and then showed that the highest free-surface wave (i.e. limiting wave, or extreme wave) in deep water would have a sharply pointed crest, enclosing a $120^\circ$ angle.  \citet{Schwartz1974Computer} carried out this expansion for deep-water wave to the order 70 and found that, as the wave height $H$ increases,  the first Fourier coefficient $a_1$ first increases until it reaches a peak value, and then decreases.    In other words,  a single $a_1$ corresponds to two different wave heights for large enough  wave height $H$.  Thus, Stokes' expansion for deep-water waves is invalid for the limiting/extreme wave height.

Then \citet{Schwartz1974Computer} used a new expansion parameter $\epsilon = H/2$ in his perturbation approach, and carried out the perturbation expansion to the 117th order  in deep water and to the 48th order in general water depths, respectively.    Utilizing the Pad\'{e} approximants and the Shanks's iterated $e_1$ transformations \citep{Shanks1954Non}, \citet{Schwartz1974Computer} successfully obtained convergent results for the ratio of water depth to wavelength $d/\lambda>0.05$.  However, his method has to rely on the extrapolation to obtain the dispersion relation for very high wave, since his perturbation series for the square of phase velocity, $c^2$, only converges well for wave height shorter than $97\%$ of the maximum.  In addition, \citet{Schwartz1974Computer} found that accurate wave profile cannot be obtained for very high waves even with the aid of Pad\'{e} approximant, so he added some standard terms to the crest to account for the remainder of the profile.  Note that \citet{Schwartz1974Computer} ascribed the failure of his perturbation method in shallow water to round-off error.

A new perturbation quantity
\begin{equation}
\epsilon=1-\frac{v_{crest}^2}{c_0^2},
\end{equation}
where $v_{crest}$ and $c_0$ are the fluid speed at the crest in the reference frame moving with the wave and the speed of waves of infinitesimal amplitude, respectively, was considered by \citet{Longuethiggins1974On}.  Using this perturbation quantity, \citet{Longuethiggins1974On} found that the series under the use of Pad\'{e}-approximants converges better than that using $\epsilon=H/2$.
Further, another expansion parameter
 \begin{equation}
\epsilon=1-\frac{v_{crest}^2 v_{trough}^2}{c^2 c_0^2}, \label{Longuet1975:parameter}
\end{equation}
where $v_{trough}$ and $c$ denote the fluid speed at the trough and the phase speed in the inertial frame, respectively, was used by \citet{Longuet1975Integral}.  The computational efficiency was drastically improved  by using this perturbation quantity.  In addition, \citet{Longuet1975Integral} proposed an alternative expansion parameter
\begin{equation}
\epsilon=1-\frac{v_{crest}^2 v_{trough}^2}{c^4}. \label{Cokelet1977:parameter}
\end{equation}
Using (\ref{Cokelet1977:parameter}) as the perturbation quantity, \citet{Cokelet1977Steep} carried out the expansion to the 120th order, and obtained convergent results for Stokes waves with $d/\lambda>0.0168$.  However, \citet{Cokelet1977Steep} pointed out that his method {\em cannot} give accurate wave profiles even in case of $d/\lambda<0.11$.  Furthermore,  \citet{Dallaston2010Accurate} reconsidered both Schwartz's \citep{Schwartz1974Computer} and Cokelet's \citep{Cokelet1977Steep} schemes,  but with exact calculations so as to void any round-off error.   However,  they found that both the series expansions of \citet{Schwartz1974Computer} and \citet{Cokelet1977Steep} actually \emph{cannot} provide precise estimates of the limiting wave properties in extremely shallow water.

Besides perturbation methods \citep{Schwartz1974Computer, Cokelet1977Steep}, a variety of numerical methods were proposed for the limiting Stokes' wave.  One common numerical method is to minimize the mean-squared error in the kinematic and dynamic free surface boundary conditions.  \citet{Chappelear1961Direct} expanded the velocity and the profile equation in Fourier series, and then used the method of least squares to determine the Fourier coefficients.  \citet{Dean1965Stream} employed an analytical stream function expression with a series of unknown coefficients to describe the waves, and then used a numerical perturbation method to determine the unknown coefficients.  Similarly, the numerical method was used by \citet{Williams1981Limiting} to minimize the error in the surface boundary conditions over a series of evenly spaced points.  However, a new crest term was supplemented to the integral equation in Williams' numerical method.   \citet{Williams1981Limiting} found that introducing this new term can greatly accelerate the convergence, i.e., the same level of accuracy can be reached by less terms of Fourier coefficients.  This method \citep{Williams1981Limiting} was a significant progress in numerics for Stokes' wave, since it is one of few methods that are both free from extrapolation and can give accurate results.  Unfortunately, this method \citep{Williams1981Limiting} still fails for the extremely shallow water $d/\lambda<0.0168$.

\citet{Rienecker1981A} used a finite Fourier series to reduce the free surface conditions to a set of nonlinear algebraic equations, and then used Newton's iteration method to solve these nonlinear equations.  By means of this method, the equations are satisfied identically at a number of points on the surface.  This method was further simplified by \citet{Fenton1988The} who numerically solved all the necessary partial derivatives.  However, \citet{Fenton1988The} found that it is sometimes still necessary first  to   solve a sequence of lower waves and then to extrapolate forward in height steps to reach the desired height.  \citet{J1979Numerical} proposed an efficient numerical scheme to solve the steep gravity wave.  They first formulated the steep gravity waves as a system of integro-differential equations, and then used the Newton's iteration technique to solve the coupled equations.  Using this numerical method, accurate results can be obtained even in the case of $d/\lambda=0.008$.  In addition, \citet{Vanden1995Computations} employed series truncation methods, which use a refinement of Davies-Tulin's approximation \citep{davies1951the, Tulin1983An}, to solve the steep gravity waves.  By means of these methods, accurate numerical results can be obtained in the cases of $d/\lambda\geq0.0168$.  It should be emphasized that these schemes are easier to implement than the boundary integral equation methods \citep{Hunter1983Accurate}.

Besides the property of limiting Stokes wave with a included $120^\circ$ angle in the crest, the non-monotonicity of the speed and energy near the limiting wave height, first found by \citet{Longuet1975Integral}, also received wide attention from researchers.  \citet{Longuet-higgins1978Theory} proposed a matching technique for gravity waves of almost extreme form, and then successfully confirmed the existence of branch-points of order $1/2$, as predicted by \citet{Grant1973the}, and of turning-points in the phase velocity as a function of wave height.  In addition, the asymptotic solution of \citet{Longuet-higgins1978Theory} indicates that there is an infinite number of turning-points in the dispersion relation, momentum and energy for the wave height very close to maximum height.  However, many methods generally only capture one or two of these turning points \citep{Chandler1993The}, but three turning points are found by \citet{Dallaston2010Accurate}.

Note that understanding the characterization of the singularity structure of the Stokes' wave, such as the locations and scalings of the singularities, is of great help in theory \citep{Tanveer137, Crew2016New}. \citet{Dyachenko2014Complex, Dyachenko2016Branch} analyzed the distance $d_c$ from the lowest singularity in the upper half-plane (i.e., the square-root branch point) to the real line which corresponds to the fluid free surface, and then suggested a power law scaling $d_c\propto (H_{max}-H)^{3/2}$.  Using this power law scaling, \citet{Dyachenko2014Complex, Dyachenko2016Branch} presented an estimate $H_{max}/\lambda\approx0.1410633$ in deep water.  Moreover, a square-root branch point is found by \citet{Lushnikov2016Structure} to be the only singularity in the physical (first) sheet of Riemann surface for non-limiting Stokes wave.  Then infinite number of square root singularities are found in the infinite number of non-physical sheets of Riemann surface after crossing a branch cut of a square root into the second and subsequently higher sheets of the Riemann surface.  Furthermore, \citet{Lushnikov2016Structure} conjectured that non-limiting Stokes wave at the leading order consists of the infinite product of nested square root singularities, and that as increasing the steepness of the Stokes wave to the extreme form, these nested square root singularities will simultaneously approach to the real line from different sheets of Riemann surface and finally form together $2/3$ power law singularity of the limiting Stokes wave.  This conjecture was well supported by high precision simulations. In addition, the slow decay of the Fourier coefficients is a challenging problem for numerical methods due to the existence of the singularities for limiting/approximately-limiting Stokes' wave.  In order to move all complex singularities away from the the free surface, \citet{Lushnikov2017New} introduced a free parameter into an auxiliary conformal mapping so as to allow finer resolution near the crest of the wave.  They found that the numerical convergence rate is dramatically improved by adapting the numerical grid near singularities.

Up to now, there are only few methods \citep{Schwartz1974Computer, Cokelet1977Steep, Williams1981Limiting, J1979Numerical} capable to solve the two-dimensional limiting (extreme) steady progressive wave in very shallow water.  Besides, almost all analytic/numerical methods fail to give accurate results (especially the wave profile) for limiting waves in extremely shallow water, such as $d/\lambda<0.005$.  In addition, most analytic/numerical methods rely on the extrapolation techniques, such as the Pad\'{e} approximant, so as to accelerate the convergence and remove singularities that limit a series radius of convergence.   So, an approach that can yield accurate results for the two-dimensional {\em limiting} (extreme) progressive gravity wave in \emph{arbitrary} water depth without using any kind of extrapolation techniques is of great value.   This is the motivation of this paper.

In this paper, the limiting Stokes' wave in \emph{arbitrary} water depth is successfully solved by an analytic approximation method, namely the homotopy analysis method (HAM) \citep{LiaoPhd, Liao1999A, Liaobook,  Liao2009, Liao2010, liaobook2, KV2008, Mastroberardino2011Homotopy, Kimiaeifar2011Application, KV2012, Duarte2015, Liao2016JFM, Zhong2017, Zhong2018Analytic, Liu2018Finite, Liu2018Mass}.  Unlike perturbation methods \citep{Schwartz1974Computer, Cokelet1977Steep}, the HAM is independent of \emph{any} small/large physical parameter.  Especially,  different from all analytic approximations, there is a so-called ``convergence-control parameter" $\hbar$ in the frame of the HAM, which has no physical meaning but provides us a convenient way to guarantee the convergence of series solutions.  
For example, perturbation techniques are invalid for large deformation of Von K\'{a}rm\'{a}n plate, a famous classic problem in solid mechanics.  However, \cite{Zhong2017, Zhong2018Analytic} successfully applied the HAM to gain convergent series solution even  
for {\em extremely}  large  deformation  of  Von K\'{a}rm\'{a}n plate.  
 It is worthwhile mentioning that some mathematical theorems of convergence have been rigorously proved in the frame of the HAM \citep{liaobook2}.  For instance, it has been proved by \cite{liaobook2} that the power series given by the HAM
\begin{equation}
u(t)=\lim_{m\rightarrow+\infty}\sum_{n=0}^{m}\mu_{0}^{m,n}(\hbar)(-t)^n, \label{ham:intro:power:series}
\end{equation}
where
\begin{equation}
 \mu_{0}^{m,n}(\hbar)=(-\hbar)^{n}\sum_{k=0}^{m-n}\binom{n-1+k}{k}(1+\hbar)^{k},
\end{equation}
 converges to $1/(1+t)$ in the intervals:
\begin{equation}
-1<t<-\frac{2}{\hbar}-1,\qquad \mbox{when} \;\;\hbar<0,
\end{equation}
and
\begin{equation}
-\frac{2}{\hbar}-1<t<-1, \qquad \mbox{when}\;\;\hbar>0,
\end{equation}
respectively. So, the power series (\ref{ham:intro:power:series}) converges to $1/(1+t)$ either in the interval $(-1,+\infty)$ if letting $\hbar<0$ impend $0$, or in the interval $(-\infty,-1)$ if letting $\hbar>0$ tend to $0$, respectively. In other words,  the power series (\ref{ham:intro:power:series}) given by the HAM  can  converge  to $1/(1+t)$ in its entire definition interval (except the singularity $t=-1$)  by properly choosing the so-called ``convergence-control parameter'' $\hbar$.   This is a good example to illustrate that the HAM can {\em dramatically} improve the convergence of a series by means of the so-called convergence control parameter.   By contrast, the traditional power series:
\begin{equation}
\frac{1}{1+t}\sim 1-t+t^{2}-t^{3}+t^{4}-\cdots
\end{equation}
only converges in the interval $(-1,1)$.   Thus, the HAM  can indeed greatly enlarge the convergence interval of solution series by means of properly choosing the so-called ``convergence-control parameter" $\hbar$.  Note that perturbation methods for many problems have been found to be a special case of the HAM when $\hbar=-1$, as illustrated by \cite{Zhong2017, Zhong2018Analytic}, and this well explains why perturbation results are often invalid in case of high nonlinearity.  In addition,  the HAM provides us great freedom to choose initial approximation so that iteration can be easily used to accelerate convergence in the frame of the HAM.  Besides,  a better initial guess can also modify the convergence of iteration, too.  More importantly,  something completely new/different have been successfully obtained by means of the HAM: the steady-state resonant waves were first predicted by the HAM in theory \citep{LIAO20111274, xu2012JFM, Liu2014JFM} and then confirmed experimentally in a laboratory \citep{Liu2015JFM}: all of these illustrate the novelty and potential of the HAM for highly nonlinear problems.

There exist two challenges for the traditional perturbation methods: (1) the series solutions diverge either when the water depth is rather small or when the wave height approaches to the peak value; (2) the computational efficiency is pretty low when the terms of Fourier coefficients are large.  For the first challenge, it is found that the convergence of series solutions of the limiting Stokes wave can be guaranteed by means of choosing a proper convergence-control parameter $\hbar$ in the frame of the HAM.   Note that, since there is a singularity exactly locating at the crest for Stokes' wave of extreme form, considering  many  enough  terms of Fourier series is inevitable if results of high precision are required.   For the second challenge, we used an iteration HAM approach to greatly accelerate convergence of {\em all} unknown Fourier coefficients.  Thus, by means of an iteration HAM approach with properly choosing convergence-control parameter $\hbar$, accurate results in \emph{arbitrary} water depth can be obtained efficiently.   More importantly, since {\em all} Fourier coefficients obtained by the HAM are convergent, accurate wave profiles in very shallow water can be presented {\em without} using any kinds of extrapolation  techniques.  Compared with the perturbation methods \citep{Schwartz1974Computer, Cokelet1977Steep}, our HAM approach is simpler, more easy-to-use and valid in almost the whole range of physical parameters, as mentioned later in this paper.   All of these demonstrate the superiority of the HAM over the perturbation method for this famous problem in fluid mechanics.

This paper is organized as follows.  Fundamental equations  are given in \S~2.   Procedures of the HAM for the limiting Stokes wave problem are presented in \S~3.  The limiting  (extreme) Stokes wave in infinite depth is considered in \S~4.   The limiting Stokes waves in finite depth are investigated in \S~5, with comparison to previous results \citep{Laitone1960The,  Fenton1972A, Schwartz1974Computer, Cokelet1977Steep, Williams1981Limiting}.  Concluding remarks are given in \S~6.

\section{Mathematical description of limiting Stokes' wave}

Consider symmetrical, two-dimensional, periodic gravity  waves propagating from right to left in a fluid with a horizontal bottom.  The propagation of waves is only under the influence of gravity.  Wave speed, $c$, is constant relative to an inertial frame.  Assume that the fluid is inviscid and incompressible and that the motion is irrotational.  Consider another reference frame moving with a wave crest.  With respect to this frame, the motion is steady.

\begin{figure}
\centering
\subfigure[]{
\begin{minipage}[b]{4.6in}
\centering
\includegraphics[width=3.5in]{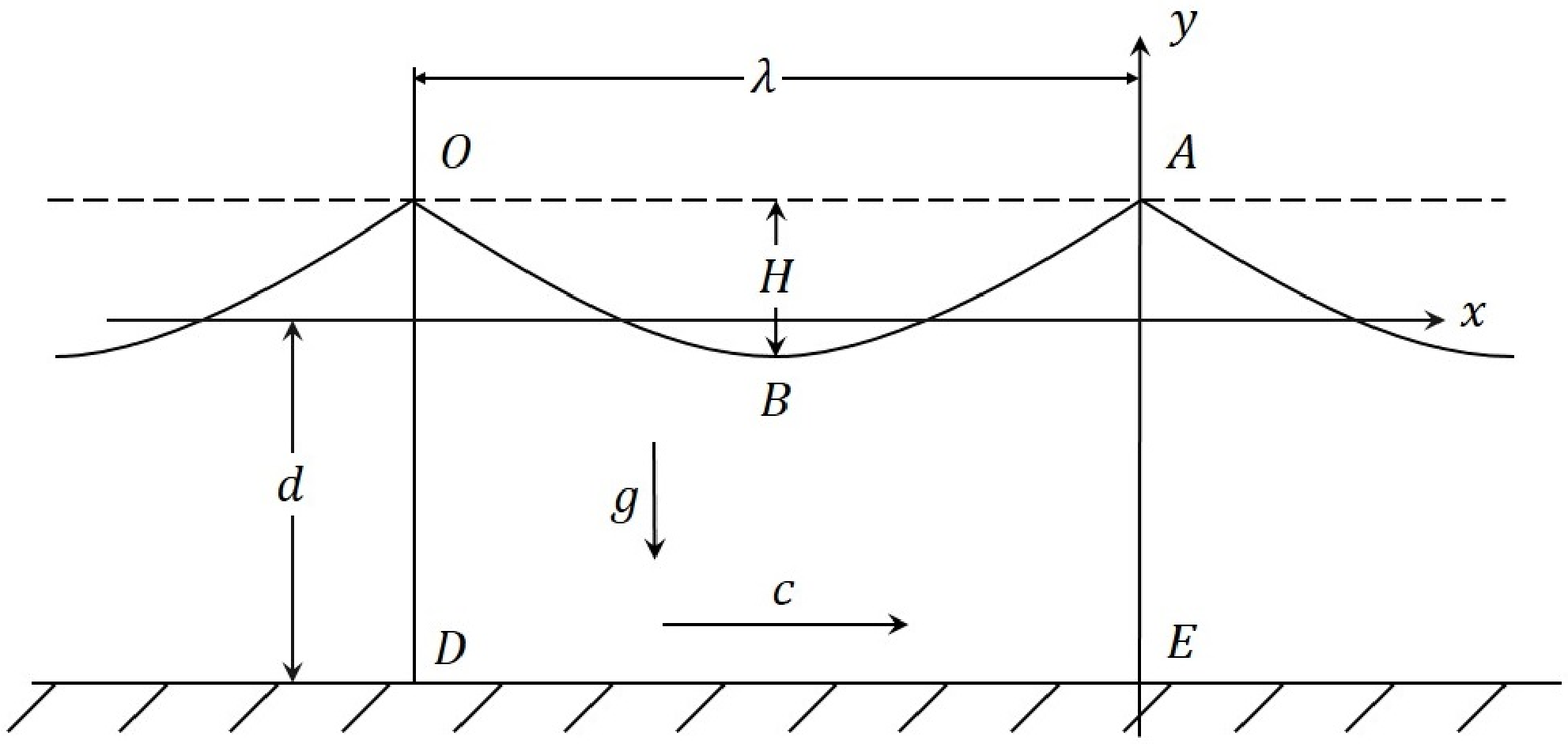}
\end{minipage}
}\\

\subfigure[]{
\begin{minipage}[b]{4.6in}
\centering
\includegraphics[width=3.5in]{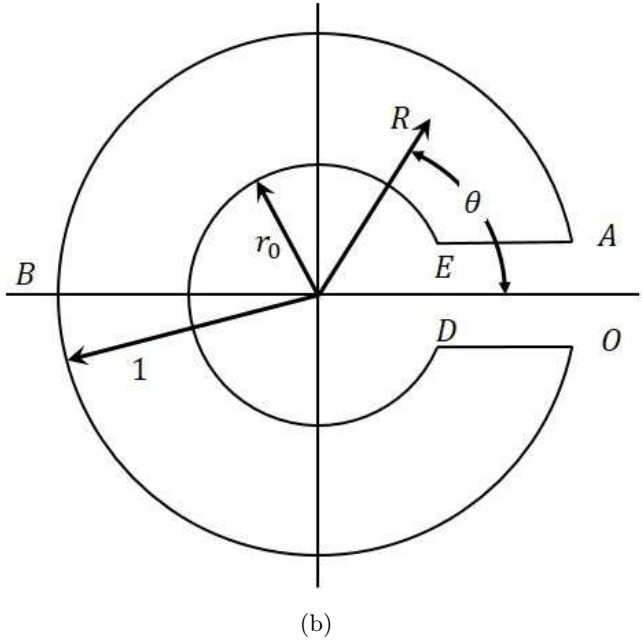}
\end{minipage}
}\caption{(a) $z$ plane, (b) $\zeta$ plane} \label{plane:trans}
\end{figure}

As shown in figure \ref{plane:trans} (a), $\lambda$, $H$, $g$ represent the wavelength, the wave height and the gravity acceleration,  respectively.  Locate the $x$ axis at a distance $d$ above the bottom.  Let the stream function $\Psi=0$ on the free surface, and $\Psi=-c \; d$ on the horizontal bottom.  The Bernoulli condition on the free surface reads
\begin{equation}
v\bar{v}+2gy=K, \quad\quad  \Psi=0,  \label{Bernoulli}
\end{equation}
where velocity $v=v_x-\mathrm{i}v_y$, the bar  denotes  the complex conjugation,  and $K$ is an unknown constant, respectively.

As shown in figure \ref{plane:trans}, we map the interior of fluid motion ``ABODEA" in the physical ``$z$" plane into an annulus ``ABODEA" in the ``$\zeta$" plane according to the transformation:
\begin{equation}
x+\mathrm{i}y=z(x,y)=z(\zeta)=\mathrm{i}\left[\mathrm{ln}\zeta+\sum_{j=1}^{+\infty}\frac{a_j}{j}\left(\zeta^j-\frac{r_0^{2j}}{\zeta^j} \right)  \right],  \label{transform:1}
\end{equation}
where $\zeta=R\mathrm{e}^{\mathrm{i}\theta}$;  $R$, $r_0$ and $\theta$ represent the radius, inner radius and argument respectively; $a_1, a_2, \cdots, a_j, \cdots$ are unknown constant coefficients to be computed.   The horizontal bottom $\Psi=-c \; d$ and the free surface $\Psi=0$ are then mapped onto the circles $R=r_0=\mathrm{e}^{-d}$ and $R=1$, respectively.  Note that $r_0=0$ and $r_0=1$  correspond  to  the cases of infinite depth and infinite shallow water,  respectively.  The complex velocity potential $w$ can be expressed as
\begin{equation}
w=\Phi+ \mathrm{i}\Psi=\mathrm{i}\; c\; \mathrm{ln}\zeta=c \; \theta+\mathrm{i} \;c\;\mathrm{ln}R,   \label{define:w}
\end{equation}
where $\Phi$ represents velocity potential.

According to (\ref{transform:1}), we have
\begin{eqnarray}
\left\{
\begin{split}
&-x=\theta+\sum_{j=1}^{+\infty}\frac{a_j}{j}\left(R^j+\frac{r_0^{2j}}{R^j}\right)\sin(j\theta) ,\\
&y=\mathrm{ln}R+\sum_{j=1}^{+\infty}\frac{a_j}{j}\left(R^j-\frac{r_0^{2j}}{R^j}\right)\cos(j\theta).
\end{split}  \label{direct:stream:line}
\right.
\end{eqnarray}
So, we have the wavelength
\begin{equation}
\lambda=x\Big{|}_{R=1, \theta=0} -x\Big{|}_{R=1, \theta=2\pi} =2\pi,
\end{equation}
and the wave steepness
\begin{equation}
\frac{H}{\lambda}=\frac{1}{2\pi}\left(y\Big{|}_{R=1, \theta=0}  -y\Big{|}_{R=1, \theta=\pi} \right) =\sum_{j=1}^{+\infty}\frac{a_j}{2j\pi }\left(1-r_0^{2j}\right)\Big{[}1-\cos(j\pi)\Big{]}.  \label{def:wave:steepness}
\end{equation}

According to (\ref{define:w}), the complex velocity $v$ reads
\begin{equation}
v=\frac{\mathrm{d}w}{\mathrm{d}z}=\frac{\mathrm{d}w}{\mathrm{d}\zeta}\; \frac{\mathrm{d}\zeta}{\mathrm{d}z}=\frac{c}{f(\zeta)},  \label{velocity}
\end{equation}
where
\begin{equation}
f(\zeta)=1+\sum_{j=1}^{+\infty}a_j\left(\zeta^j+\frac{r_0^{2j}}{\zeta^j}\right).  \label{def:f}
\end{equation}
Note that the velocity at the crest is zero for the highest wave.  Using this restriction condition, Eq.~(\ref{Bernoulli}) becomes
\begin{equation}
v\bar{v}+2g\int_0^{\theta}\mathrm{Im}\left[\frac{\mathrm{d}z}{\mathrm{d}\zeta} \; \frac{\mathrm{d}\zeta}{\mathrm{d}\theta}\right]\mathrm{d}\theta=0,\quad\quad \mbox{when $\Psi=0$}. \label{free:1}
\end{equation}
Substituting (\ref{transform:1}), (\ref{velocity}), (\ref{def:f}) into (\ref{free:1}), we have the nonlinear algebraic equation
\begin{equation}
\frac{2g}{c^2}\;f\;\bar{f}\;\int_{0}^{\theta}\textrm{Im}\big{[}f \big{]}\textrm{d}\theta-1=0, \quad\quad \mbox{at $R=1$}.\label{free:end}
\end{equation}

Theoretically, $a_1$, $a_2$, $\cdots$, $a_j$, $\cdots$ need to be all reserved to identically satisfy the equation (\ref{free:end}). However,  we can  only consider limited terms in practice.  Thus, let us consider here  the first $r$ Fourier coefficients $a_1$, $a_2$, $\cdots$, $a_r$, i.e., $f$ is approximated by
\begin{equation}
f(\zeta)\approx a_0+\sum_{j=1}^{r}a_j\left(\zeta^j+\frac{r_0^{2j}}{\zeta^j}\right),\quad\quad  a_0=1. \label{f:expression}
\end{equation}
Substituting (\ref{f:expression}) into (\ref{free:end}) and then equating the coefficients of $\cos(k \theta)$, where $k=0,1,2,\cdots,r$, we obtain the following $(r+1)$ algebraic equations \footnote{Detailed derivation is shown in Appendix.}:
\begin{equation}
c^2 =g\left(2j_0 h_0+\sum_{n=1}^{r}j_n h_n \right) ,\label{origin:equation:B}
\end{equation}
and
\begin{eqnarray}
&&{\cal N}_k[a_1,a_2,\cdots,a_r]\nonumber\\
&=&  j_0 h_k + j_k h_0 +\frac{1}{2} \left(\sum_{n=1}^{k-1}j_n h_{k-n}+\sum_{n=1}^{r-k}j_n h_{n+k}+\sum_{n=1}^{r}j_{n+k} h_n \right)=0, \label{origin:equation:A}
\end{eqnarray}
where ${\cal N}_k$ ($k=1,2,\cdots,r$) denotes a nonlinear operator,  with the following definitions:
\begin{eqnarray}
\left\{
\begin{split}
h_0&=\sum_{n_1=1}^{r}\frac{a_{n_1}\left(1-r_0^{2 n_1} \right)}{n_1},\quad\quad \\
h_n&=-\frac{a_n\left(1-r_0^{2 n} \right)}{n}\quad \mbox{ when $1\leq n\leq r$},\\
j_0&=1+\sum_{n_1=1}^{r} a_{n_1}^2\left(1+r_0^{4 n_1} \right),\\
j_n&=2\left[\sum_{n_1=0}^{r-n}\left(1+r_0^{2n+4 n_1} \right)a_{n_1}a_{n_1+n}+\sum_{n_1=1}^{n-1}r_0^{2n-2n_1}a_{n_1}a_{n-n_1} \right]\\
&\quad \mbox{ when $1\leq n \leq r$},\\
j_n&=2\sum_{n_1=n-r}^{r}r_0^{2n-2n_1}a_{n_1}a_{n-n_1}\quad \mbox{ when $ r< n \leq 2r$}.\\
\end{split}  \label{def:j:h}
\right.
\end{eqnarray}
Then, the next step is to solve the nonlinear algebraic equations (\ref{origin:equation:A}) for the $r$ unknown constant Fourier coefficients $a_1$, $a_2$, $\cdots$, $a_r$.  Thereafter,  the wave speed $c$ can be directly given by (\ref{origin:equation:B}).

\renewcommand{\theequation}{3.\arabic{equation}}

\setcounter{equation}{0}
\section{The mathematical approach based on the HAM}
Let $a_{j,0}$ denote the initial guess of $a_j$ $(j=1,2,\cdots,r)$, $\hbar$  a non-zero auxiliary parameter (called the convergence-control parameter), and $q\in [0,1]$ the embedding parameter for a homotopy, respectively.   First of all, we construct a family of equations
\begin{equation}
(1-q)\Big{[}\Omega_k(q)-a_{k,0}\Big{]}=\hbar \; q  \;  {\cal N}_k \Big{[} \Omega_1(q), \Omega_2(q), \cdots,\Omega_r(q)\Big{]}, \quad\quad k=1,2,\cdots,r, \label{zero:deformation}
\end{equation}
where the nonlinear operators ${\cal N}_1, {\cal N}_2, \cdots, {\cal N}_r$ are defined by
(\ref{origin:equation:A}),  and the unknown functions  $\Omega_1(q)$, $\Omega_2(q)$, $\cdots$, $\Omega_r(q)$ correspond to the unknown constant Fourier coefficients  $a_1$, $a_2$, $\cdots$, $a_r$, respectively, and $a_{k,0}$ is the initial guess of $a_k$.  Note that, in the frame of the HAM, we have great freedom to choose the initial guess $a_{k,0}$ so that an iteration approach can be proposed based on this kind of freedom to greatly accelerate convergence, as mentioned later.  More importantly, the so-called convergence-control parameter $\hbar$ can provide us a simple way to guarantee the convergence of solution series, as shown below. 

Obviously, when $q=0$, Eq.~(\ref{zero:deformation}) has the solution
\begin{equation}
    \Omega_k(0)=a_{k,0},\quad\quad  k=1,2,\cdots,r. \label{initial}
\end{equation}
 When $q=1$, Eq.~(\ref{zero:deformation}) is equivalent to the original equation (\ref{origin:equation:A}), provided
\begin{equation}
    \Omega_k(1)=a_{k},\quad\quad  k=1,2,\cdots,r. \label{similar}
\end{equation}
Therefore, as $q$ increases from $0$ to $1$, the function $\Omega_j(q)$ varies (deforms) continuously from the known initial guess  $a_{j,0}$ to the unknown constant Fourier coefficient $a_j$, where $j=1,2,\cdots,r$.   In the frame of the HAM,  equation (\ref{zero:deformation}) is called the zeroth-order deformation equations.  Obviously, according to (\ref{initial}), we have the Maclaurin series
\begin{equation}
\Omega_n(q)=a_{n,0}+\sum_{k=1}^{+\infty}a_{n,k}\; q^k,\quad\quad n=1,2,\cdots,r,
\label{power:series}
\end{equation}
where
\begin{equation}
   a_{n,k}={\cal D}_{k}\big{[}\Omega_n(q)\big{]}, \quad\quad n=1,2,\cdots,r, \label{homotopy:decree}
\end{equation}
in which
\begin{equation}
   {\cal D}_{k}\big{[}f\big{]}=\frac{1}{k!}\frac{\partial^{k}f}{\partial q^{k}}\bigg{|}_{q=0} \label{Dm}
\end{equation}
is called the $k$th-order homotopy-derivative of $f$.   Note that,  according to (\ref{zero:deformation}),  $\Omega_n(q)$ and its series (\ref{power:series}) are dependent upon the so-called convergence-control parameter $\hbar$.    Assuming that  $\hbar$ is properly chosen so that the Maclaurin series  (\ref{power:series}) exists and converges at $q=1$,    then  according to (\ref{similar}),  we have the so-called homotopy-series solutions
\begin{equation}
    a_n=\sum_{k=0}^{+\infty}a_{n,k},\quad\quad n=1,2,\cdots,r. \label{homotopy:series}
\end{equation}

Substituting (\ref{power:series}) into the zeroth-order deformation equations (\ref{zero:deformation}) and then equating the like-power of $q$, we have the so-called $m$th-order deformation equations
\begin{eqnarray}
a_{k,m}-\chi_m a_{k,m-1}=\hbar\; {\cal D}_{m-1}\big{[}{\cal N}_k\big{]},\quad\quad k=1,2,\cdots,r,
\end{eqnarray}
where
\begin{eqnarray}
&&\mbox{}{\cal D}_i\big{[}{\cal N}_k\big{]} \nonumber\\
&=&\sum_{n_2=0}^{i}\Bigg{\{}-\left[\frac{a_{k,i-n_2}\left( 1-r_0^{2k}\right)}{k} \right]  \nonumber\\
&&\mbox{}\times\left[1-\chi_{n_2+1}+ \sum_{n_1=1}^{r}\sum_{n_3=0}^{n_2}\left( 1+r_0^{4n_1}\right)a_{n_1,n_3}a_{n_1,n_2-n_3}   \right]\nonumber\\
&&\mbox{}+2\left[\sum_{n_1=1}^{r}\frac{a_{n_1,i-n_2}\left( 1-r_0^{2n_1}\right)}{n_1} \right] \Bigg{[}\sum_{n_1=1}^{k-1}\sum_{n_3=0}^{n_2}r_0^{2k-2n_1}a_{n_1,n_3}a_{k-n_1,n_2-n_3}\nonumber\\
&&\mbox{}+\sum_{n_1=0}^{r-k}\sum_{n_3=0}^{n_2}\left(1+r_0^{2k+4n_1}\right)a_{n_1,n_3}a_{n_1+k,n_2-n_3}\Bigg{]}\nonumber\\
&&\mbox{}-\sum_{n=1}^{k-1}\left[\frac{a_{k-n,i-n_2}\left( 1-r_0^{2k-2n}\right)}{k-n}\right]\Bigg{[}\sum_{n_1=1}^{n-1}\sum_{n_3=0}^{n_2}r_0^{2n-2n_1}a_{n_1,n_3}a_{n-n_1,n_2-n_3}\nonumber\\
&&\mbox{}+\sum_{n_1=0}^{r-n}\sum_{n_3=0}^{n_2}\left(1+r_0^{2n+4n_1}\right)a_{n_1,n_3}a_{n_1+n,n_2-n_3}\Bigg{]} \nonumber\\
&&\mbox{}-\sum_{n=1}^{r-k}\left[\frac{a_{k+n,i-n_2}\left( 1-r_0^{2k+2n}\right)}{k+n}\right]\Bigg{[}\sum_{n_1=1}^{n-1}\sum_{n_3=0}^{n_2}r_0^{2n-2n_1}a_{n_1,n_3}a_{n-n_1,n_2-n_3}\nonumber\\
&&\mbox{}+\sum_{n_1=0}^{r-n}\sum_{n_3=0}^{n_2}\left(1+r_0^{2n+4n_1}\right)a_{n_1,n_3}a_{n_1+n,n_2-n_3}\Bigg{]}\nonumber\\
&&\mbox{}-\sum_{n=1}^{r-k}\left[\frac{a_{n,i-n_2}\left( 1-r_0^{2n}\right)}{n}\right]\Bigg{[}\sum_{n_1=1}^{n+k-1}\sum_{n_3=0}^{n_2}r_0^{2n+2k-2n_1}a_{n_1,n_3}a_{n+k-n_1,n_2-n_3}\nonumber\\
&&\mbox{}+\sum_{n_1=0}^{r-n-k}\sum_{n_3=0}^{n_2}\left(1+r_0^{2n+2k+4n_1}\right)a_{n_1,n_3}a_{n_1+n+k,n_2-n_3}\Bigg{]}\nonumber\\
&&\mbox{}-\sum_{n=r-k+1}^{r}\left[\frac{a_{n,i-n_2}\left( 1-r_0^{2n}\right)}{n}\right]  \nonumber \\ &&\mbox{}\times\Bigg{[}\sum_{n_1=n+k-r}^{r}\sum_{n_3=0}^{n_2}r_0^{2n+2k-2n_1}a_{n_1,n_3}a_{n+k-n_1,n_2-n_3} \Bigg{]} \Bigg{\}}, \label{high:deformation}
\end{eqnarray}
in which
\begin{equation*}
a_{0,0}=1,\quad\quad\quad a_{0,k}=0\quad\mbox {when \; $k\geq 1$},
\end{equation*}
and
\begin{equation}
\chi_k =\left\{
\begin{array}{cc}
0 & \mbox{when $k\leq1$}, \\
1 & \mbox{when $k >1$.}
\end{array}
\right. \label{def:chi}
\end{equation}

Note that, in the frame of the HAM, we have great freedom to choose the initial guesses $a_{1,0}$, $a_{2,0}$, $\cdots$, $a_{r,0}$.  So,  we  can simply choose
\begin{eqnarray}
 a_{k,0} = \frac{1}{k}, \quad\quad k=1,2\cdots,r. \label{initial:guesses}
\end{eqnarray}
Then $a_{1,k}$, $a_{2,k}$, $\cdots$, $a_{r,k}$ can be obtained by (\ref{high:deformation}) step by step, starting from $k=1$. The $n$th-order homotopy approximations of $a_1$, $a_2$, $\cdots$, $a_r$ read
\begin{equation}
   \tilde{\Omega}_{i,n}=\sum_{k=0}^{n}a_{i,k},\quad\quad i=1,2\cdots,r.   \label{sum}
\end{equation}
Once $a_1$, $a_2$, $\cdots$, $a_r$ are determined, the wave speed $c$ can be given by (\ref{origin:equation:B}).

In order to characterize the global error of our HAM approximation, we define the following squared residual error
\begin{equation}
   {\cal E}=\sum_{i=1}^{r}\left({\cal N}_{i}\left[\tilde{\Omega}_1, \tilde{\Omega}_2, \cdots, \tilde{\Omega}_r\right]\right)^{2},  \label{homotopy:error}
\end{equation}
where the nonlinear operators  ${\cal N}_1$, ${\cal N}_2$, $\cdots$, ${\cal N}_r$ are defined by (\ref{origin:equation:A}).  Obviously,  the  smaller  the ${\cal E}$,  the  more  accurate  the  HAM  approximation  (\ref{sum}).  Besides, it has been proved \citep{Liaobook, liaobook2} in general that a  homotopy-series converges to solution of original equations as long as all squared residual errors tend to zero.   So,  it is enough to check the squared residual error (\ref{homotopy:error}) only.

\renewcommand{\theequation}{4.\arabic{equation}}
\section{The limiting Stokes' wave in infinite depth}

To show the validity of our HAM approach mentioned above,  we first of all give convergent series solution of the limiting (extreme) Stokes' wave in infinite depth.

According to \citet{Liaobook}, the convergence of the homotopy-series solutions can be greatly accelerated by introducing the iteration technique, which uses the $n$th-order homotopy-approximation  $\tilde{\Omega}_{1,n}$, $\tilde{\Omega}_{2,n}$, $\cdots$, $\tilde{\Omega}_{r,n}$ as new initial guesses $a_{1,0}$, $a_{2,0}$, $\cdots$, $a_{r,0}$ for the next iteration, say,
$a_{1,0}=\tilde{\Omega}_{1,n}$, $a_{2,0}=\tilde{\Omega}_{2,n}$, $\cdots$, $a_{r,0}=\tilde{\Omega}_{r,n}$.   This provides us the $n$th-order iteration of the HAM.
According to our computation, both the HAM approach without iteration and the HAM-based iteration approach can yield convergent results, but the efficiency of the HAM-based iteration approach is much higher.   In particular, the first-order HAM-based iteration approach has the highest efficiency.  So, we use the first-order HAM-based iteration approach in all cases of this paper, if not specially mentioned.

Table \ref{table:different:m} presents the results in the case of $r_0=0$, corresponding to infinite depth of water, given by the convergence-control parameter $\hbar=-0.2$, $r = 100$ (i.e. one hundred  truncated terms of Fourier series)  and the initial guess (\ref{initial:guesses}).   Note that the squared residual error $\cal E$ defined by (\ref{homotopy:error}) quickly decreases to the tiny level $10^{-17}$.  This illustrates that {\em all} Fourier coefficients $a_1$, $a_2$, $\cdots$, $a_{100}$, given by our HAM approach, are convergent.

\begin{table}
\tabcolsep 0pt
\begin{center}
\def\temptablewidth{0.9\textwidth}
\begin{tabular*}{\temptablewidth}{@{\extracolsep{\fill}}cccc}
$m$, iteration times  &${\cal E}$ &   $H/\lambda$    & $(g\lambda)/(2\pi c^2)$   \\
20  & $1\times10^{-2}$  & $0.10623$    &1.0573    \\
50  & $3\times10^{-3}$  & $0.15083$    &0.8153       \\
100 &   $5\times10^{-5}$ &  $0.13846$      &0.8494       \\
200 &  $3\times10^{-9}$ &   $0.13974$    &0.8422          \\
300 &  $3\times10^{-13}$ &   $0.13973$    &0.8422          \\
400 &  $1\times10^{-17}$ &   $0.13973$    &0.8422          \\
\end{tabular*}
\caption{The squared residual error $\cal E$, wave steepness $H/\lambda$ and wave speed parameter $(g\lambda)/\left(2\pi c^2\right)$ versus iteration times in the case of $r_0=0$, given by the first-order HAM-based iteration approach using $c_0=-0.2$, $r=100$, and the initial guess  (\ref{initial:guesses}).}   \label{table:different:m}
 \end{center}
 \end{table}

\begin{figure}
  \centerline{\includegraphics[width=3.5in]{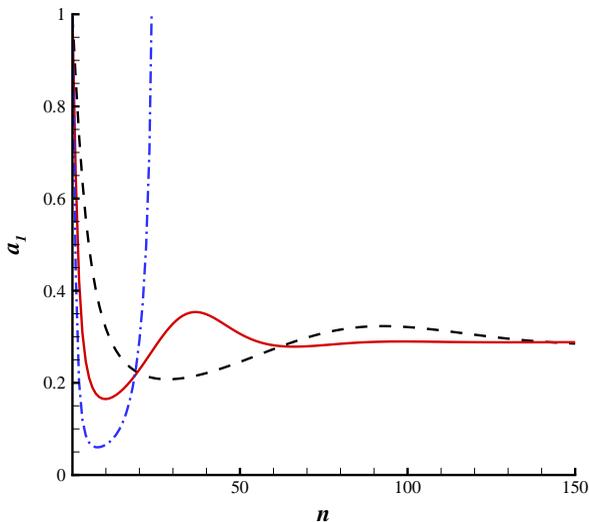}}
  \caption{The first Fourier coefficient, $a_1$, versus iteration times, $n$, in the case of $r_0=0$, given by the first-order HAM-based iteration approach using $r=100$ and the convergence-control parameter $\hbar=-0.1, -0.25, -0.4$ respectively.  -- -- --, $\hbar=-0.1$; ---$\!$---, $\hbar=-0.25$; -- $\cdot$ --, $\hbar=-0.4$.}
\label{figure:c0:a1}
\end{figure}

Figure \ref{figure:c0:a1} shows the homotopy-approximation of the first Fourier coefficient, $a_1$, versus iteration times in the case of $r_0=0$,  given by $r=100$ and the convergence-control parameter $\hbar=-0.4,-0.25,-0.1$, respectively.  Note that $\hbar=-0.4$ leads to divergence of iteration, $\hbar=-0.25$ corresponds to a quickly convergent iteration, but $\hbar=-0.1$ a slowly convergent iteration, respectively.  Obviously, the optimal value of $\hbar$ corresponds to the fastest convergence, as pointed out by \cite{Liao2010}.   It is found that  convergent results can be obtained by our iteration HAM approach with \emph{arbitrary} values of $\hbar\in[-0.27,0)$.   So, the convergence-control parameter $\hbar$ indeed provides us a simple way to guarantee convergence and to accelerate convergence.   This clearly illustrates the important role of the convergence-control parameter $\hbar$ in the frame of the HAM.

Table \ref{table:different0:r} presents the convergent results in the case of $r_0=0$, given by different $r$.  Note that the steepness of the limiting wave in infinite water depth tends to a fixed value $H/\lambda=0.14108$ when $r$ is large enough, say, $r>5000$.   This is reasonable, since the precision of our results is controlled by $r$, i.e. the truncated number of the Fourier series (\ref{f:expression}).
 Note that \citet{Schwartz1974Computer} gave  $H_{max}/\lambda=0.14118$  but \citet{Dyachenko2016Branch} gave $H_{max}/\lambda=0.141058$  for limiting wave in deep water, with $0.071\%$ and  $0.016\%$ relative errors compared to our results, respectively.   It should be emphasized that, by means of the HAM, convergent results of {\em all} Fourier coefficients $a_j$ can be obtained.  This distinguishes the HAM from other methods.

  \begin{table}
 \tabcolsep 0pt
\begin{center}
\def\temptablewidth{0.9\textwidth}
\begin{tabular*}{\temptablewidth}{@{\extracolsep{\fill}}ccc}
$r$  &  $H/\lambda$    & $(g\lambda)/(2\pi c^2)$\\
50 &   0.13926& 0.8391            \\
500 &    0.14085& 0.8397         \\
1000 &    0.14102& 0.8388          \\
2000&    0.14107& 0.8383         \\
3000&  0.14108& 0.8382         \\
4000&  0.14109& 0.8382         \\
5000&  0.14108& 0.8381         \\
6000&  0.14108& 0.8381         \\
\end{tabular*}
\caption{Wave steepness $H/\lambda$ and wave speed parameter $(g\lambda)/\left(2\pi c^2\right)$ versus truncated terms $r$ in the case of $r_0=0$ (in infinite depth), given by the first-order HAM-based iteration approach using $c_0=-0.2$.}   \label{table:different0:r}
 \end{center}
 \end{table}

 It is found that the high-order Fourier coefficients $a_j$ drop rather slowly (e.g., $a_{1}=0.29223$, $a_{100}=0.01576$, $a_{500}=0.005415$, $a_{1000}=0.003739$, $a_{3000}=0.002905$, $a_{5000}=0.002759$).   We have $H/\lambda=0.14085$ even by means of $r=500$, and have the more accurate result $H/\lambda=0.14108$ by $r>5000$.  All of these indicate that $f$ defined by (\ref{def:f}) converges pretty slowly indeed.  However, it should be emphasized that,  whatever $r$ we choose, convergent values of {\em all} Fourier coefficients $a_j$ can be  directly  obtained by our iteration HAM  approach {\em without} using any extrapolation and Pad\'{e} approximant techniques.

\begin{figure}
  \centerline{\includegraphics[width=3.5in]{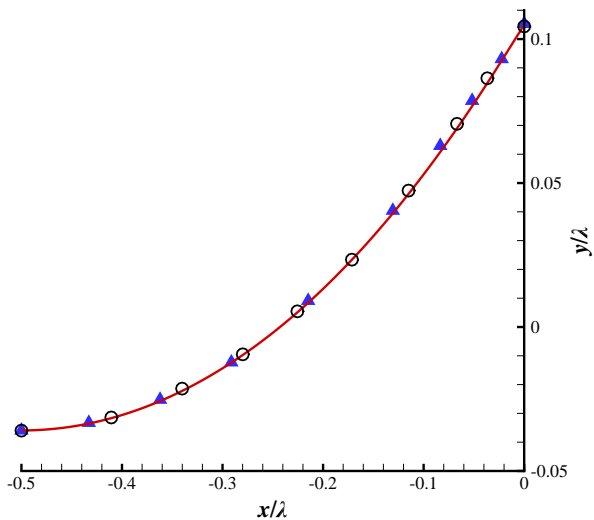}}
  \caption{Wave profiles in the case of $r_0=0$.  \protect
    ---$\!$---, given by the first-order HAM-based iteration approach; $\circ$, given by the numerical approach of  \citet{Dyachenko2016Branch};  $\blacktriangle$, given by Schwartz's perturbation method with the aid of a so-called series completion method \citep{Schwartz1974Computer}.}
\label{comparison:HAM:Schwartz}
\end{figure}

Figure~\ref{comparison:HAM:Schwartz} shows the comparison of limiting wave profiles given by \cite{Schwartz1974Computer}, \citet{Dyachenko2016Branch} and  our HAM approach.  The agreement between them is satisfactory. This indicates the validity of our HAM-based approach.   Our limiting wave profile has a sharply pointed crest with an enclosing angle $119.3^\circ$, which is very close to the theoretical value $120^\circ$.  However, \citet{Schwartz1974Computer}  mentioned that ``while the method of P\'{a}de fractions yields accurate profiles for wave heights somewhat short of the maximum,  it is {\em insufficient} for the description of very high waves'', and that ``P\'{a}de fractions do {\em not} converge well in the immediate neighbourhood of branch-points; moreover, only the first few coefficients $a_j$, can be determined with acceptable accuracy''.  Thus, \cite{Schwartz1974Computer}  had to use the so-called ``series completion method'' to gain a satisfactory wave profile.   By contrast, using the HAM, convergent results of {\em all} Fourier coefficients $a_j$ $(j=1,2,3,\cdots,r)$ and the convergent wave profile for the limiting wave can be obtained {\em without} using any extrapolation techniques such as Pad\'{e} technique, the series completion method and so on.   This illustrates that the HAM-based approach is superior to perturbation methods \citep{Schwartz1974Computer, Cokelet1977Steep}.   This is mainly because, unlike perturbation methods, the HAM provides us a convenient way (through the so-called convergence-control parameter $\hbar$) to guarantee the convergence of solution series.

\renewcommand{\theequation}{5.\arabic{equation}}

\setcounter{equation}{0}

\section{The limiting Stokes' wave in finite depth}

Note that $r_0=0$ and $r_0=1$ correspond to the  case of infinite depth and  infinite shallow water, respectively.   Without loss of generality, let us first consider the case of $r_0=0.05$.    In the frame of the HAM, we have great freedom to choose the initial guesses of $a_1$, $a_2$, $\cdots$, $a_r$.      Considering the continuous variation of $a_1$, $a_2$, $\cdots$, $a_r$ as  $r_0$  increases from 0 to 1, the convergent values of $a_1$, $a_2$, $\cdots$, $a_{5000}$ in the case of $r_0=0$, obviously, are much better than (\ref{initial:guesses}) as the initial guess for the case of $r_0=0.05$.  In other words, if we have obtained the convergent results of $a_1$, $a_2$, $\cdots$, $a_{5000}$ in the case of $r_0=0$, then it is better to take
\begin{eqnarray}
a_{k,0} =\left\{
\begin{array}{ll}
 a_{k}  \quad\quad\;\;&\mbox{when $1\leq k\leq5000$}, \\[2pt]
 a_{5000}  \quad\quad&\mbox{when $k > 5000$,}
\end{array}
\right. \label{initial:substitute}
\end{eqnarray}
as the initial guesses of $a_1$, $a_2$, $\cdots$, $a_{r}$ in the case of $r_0=0.05$.

It is found that, in the case of $r_0=0.05$, the optimal convergence-control parameter $\hbar$ is about $-0.2$ if the initial guess  (\ref{initial:guesses}) for $r_0=0$  is taken,  and  400 times iteration  is  required  to gain convergent results $H/\lambda=0.14026$ and $(g\lambda)/(2\pi c^2)=0.8421$, as shown in Table~\ref{table:r1:005:bad:initial}.  However, if we take the initial guess (\ref{initial:substitute}), the optimal convergent-control parameter $\hbar$ becomes $-1.2$, and we obtain the same convergent results  $H/\lambda=0.14026$ and $(g\lambda)/(2\pi c^2)=0.8421$ by just thirty times iteration, as shown in Table~\ref{table:r1:005:good:initial}.    Thus, the computational efficiency by means of the initial guess (\ref{initial:substitute}) is approximately 13 times higher than that by (\ref{initial:guesses}).   This illustrates that our iteration HAM approach with the optimal convergence-control parameter $\hbar$ can indeed greatly accelerate the convergence.

 \begin{table}
\tabcolsep 0pt
\begin{center}
\def\temptablewidth{0.9\textwidth}
\begin{tabular*}{\temptablewidth}{@{\extracolsep{\fill}}ccc}
$m$, iteration times  &   $H/\lambda$    & $(g\lambda)/(2\pi c^2)$   \\
10   &    $0.18670$    &0.5663    \\
50  &   $0.13681$    &0.9099      \\
100  &  $0.13976$    &0.8455       \\
200  &   $0.14033$    &0.8416         \\
300 &  $0.14023$    &0.8421      \\
400 &   $0.14026$    &0.8421         \\
500 &   $0.14026$    &0.8421         \\
\end{tabular*}
\caption{Wave steepness $H/\lambda$ and wave speed parameter $(g\lambda)/\left(2\pi c^2\right)$ versus iteration times, $m$, in the case of $r_0=0.05$, given by the first-order HAM-based iteration approach using the convergence-control parameter $c_0=-0.2$, the truncated terms $r=5500$ and the initial guess (\ref{initial:guesses}).}   \label{table:r1:005:bad:initial}
 \end{center}
 \end{table}

 \begin{table}
\tabcolsep 0pt
\begin{center}
\def\temptablewidth{0.9\textwidth}
\begin{tabular*}{\temptablewidth}{@{\extracolsep{\fill}}ccc}
$m$, iteration times  &   $H/\lambda$   & $(g\lambda)/(2\pi c^2)$   \\
10   &    $0.14018$    &0.8423    \\
20  &   $0.14024$    &0.8422       \\
30  &  $0.14026$      &0.8421       \\
40  &   $0.14026$    &0.8421          \\
50 &   $0.14026$    &0.8421          \\
\end{tabular*}
\caption{Wave steepness $H/\lambda$ and wave speed parameter $(g\lambda)/\left(2\pi c^2\right)$ versus iteration times, $m$, in the case of $r_0=0.05$, given by the first-order HAM-based iteration approach using the convergence-control parameter $c_0=-1.2$, the truncated terms $r=5500$ and the initial guesses (\ref{initial:substitute}).}   \label{table:r1:005:good:initial}
 \end{center}
 \end{table}

Similarly,   the convergent results in \emph{arbitrary} water depth are successfully obtained by means of the above-mentioned strategy,  as shown in Table~\ref{table:different:r1}.  Note that \cite{Liao2010} suggested a general  approach to gain an optimal convergence-control parameter in the frame of the HAM.   According to our computation, the interval of $\hbar$,  which guarantees the convergence  of iteration, becomes larger with the increase of $r_0$.   It is found that, in the case of  $ 0.05k < r_0  \leq 0.05(k+1)$, where $0 \leq k \leq 19$ is a natural number,  the corresponding optimal  convergence-control parameter $\hbar$ can be expressed by the following empirical formula
\begin{equation}
\hbar=-1.2-\frac{k^3}{2000},\quad\quad \mbox{ $0\leq k\leq 19$},
\end{equation}
if we use the known convergent Fourier coefficients $a_j$ in the case of $r_0 = 0.05k$ as the initial guess.
Note that a convergence-control parameter $\hbar$ closer to 0 represents a slower convergence of solutions, i.e., a lower efficiency of computation, as shown in Figure~\ref{figure:c0:a1}.  Thus,  the  convergence-control parameter $\hbar$  provides us a convenient way {\em not only} to guarantee the convergence of series solutions {\em but also} to improve the computational efficiency.   It is found that, for all cases considered in Table~\ref{table:different:r1},  a few hundred  times of  iteration are enough to gain convergent results of {\em all} Fourier coefficients $a_j$.

In case of extremely shallow water,  a huge number of Fourier coefficients are needed to present the limiting wave with the sharp crest.   Table \ref{table:different99:r} presents the convergent results given by different values of $r$ in the case of $r_0=0.99$.  It is found that $r=50000$ can give the fixed results $H/\lambda = 1.3281\times 10^{-3}$ and  $(g\lambda)/(2\pi c^2)=60.175$ in the case of $r_0=0.99$.   This indicates that our iteration HAM approach can indeed give convergent results of the limiting Stokes' waves even in extremely shallow water.  Note that, in case of $r = 50000$, we must solve a set of 50000 coupled,  highly nonlinear algebraic  equations!  Fortunately, this is possible nowadays by means of a supercomputer such as TH-2 at National Supercomputer Centre in Guangzhou, China.  Finally, it should be emphasized  that all of these convergent results are obtained directly, say, {\em without} using any extrapolation and Pad\'{e} approximant techniques.

 \begin{table}
\begin{center}
\def\temptablewidth{1\textwidth}
\begin{tabular*}{\temptablewidth}{@{\extracolsep{\fill}}cccccc}
$r_0$  & $r$ & $d/\lambda$ & $H/\lambda$  & $H/d$  & $(g\lambda)/(2\pi c^2)$  \\
0  &    5000&      $\infty$&                 $1.4108\times 10^{-1}$ & 0 & 0.8381    \\
0.05 & 5500& $4.77\times10^{-1}$&    $1.4026\times 10^{-1}$ & 0.2942 & 0.8421    \\
0.10 &  6000& $3.66\times10^{-1}$&  $1.3782\times 10^{-1}$ & 0.3761& 0.8540       \\
0.15  &6500&  $3.02\times10^{-1}$&  $1.3386\times 10^{-1}$ & 0.4433& 0.8739     \\
0.20 & 7000&  $2.56\times10^{-1}$& $1.2851\times 10^{-1}$ & 0.5017& 0.9022         \\
0.25 & 7500&  $2.21\times10^{-1}$&  $1.2197\times 10^{-1}$ & 0.5528& 0.9395     \\
0.30 & 8000&  $1.92\times10^{-1}$&  $1.1446\times 10^{-1}$ & 0.5973& 0.9864       \\
0.35  &8500&  $1.67\times10^{-1}$&  $1.0618\times 10^{-1}$ & 0.6355& 1.0442     \\
0.40&  9000&  $1.46\times10^{-1}$& $9.7388\times 10^{-2}$  & 0.6678 & 1.1145     \\
0.45  & 9500&  $1.27\times10^{-1}$&   $8.8289\times 10^{-2}$ & 0.6947& 1.2001   \\
0.50&  10000& $1.10\times10^{-1}$&  $7.9084\times 10^{-2}$ & 0.7169& 1.3048        \\
0.55 & 10500& $9.51\times10^{-2}$&  $6.9943\times 10^{-2}$ & 0.7351& 1.4344       \\
0.60 & 11000&  $8.13\times10^{-2}$&   $6.0995\times 10^{-2}$ & 0.7502& 1.5977     \\
0.65 & 11500&  $6.86\times10^{-2}$&  $5.2327\times 10^{-2}$ & 0.7632& 1.8091      \\
0.70  & 12000&  $5.68\times10^{-2}$&  $4.3983\times 10^{-2}$ & 0.7748& 2.0922  \\
0.75  & 14000&  $4.58\times10^{-2}$&   $3.5968\times 10^{-2}$ & 0.7856& 2.4898  \\
0.80  & 16000&  $3.55\times10^{-2}$&  $2.8263\times 10^{-2}$ & 0.7958& 3.0876  \\
0.85  & 18000& $2.59\times10^{-2}$&   $2.0840\times 10^{-2}$ & 0.8057& 4.0856  \\
0.90  & 22000& $1.68\times10^{-2}$&   $1.3670\times 10^{-2}$ & 0.8152& 6.0838  \\
0.95  &28000&  $8.16\times10^{-3}$&  $6.7292\times 10^{-3}$ & 0.8243& 12.084  \\
0.97  & 37000& $4.85\times10^{-3}$&  $4.0128\times 10^{-3}$ & 0.8278& 20.087  \\
0.99  & 50000& $1.60\times10^{-3}$&   $1.3281\times 10^{-3}$ & 0.8303& 60.175  \\
\end{tabular*}
\caption{Results for a variety of water depths, given by the first-order HAM-based iteration approach.}   \label{table:different:r1}
 \end{center}
 \end{table}

\begin{figure}
  \centerline{\includegraphics[width=3.5in]{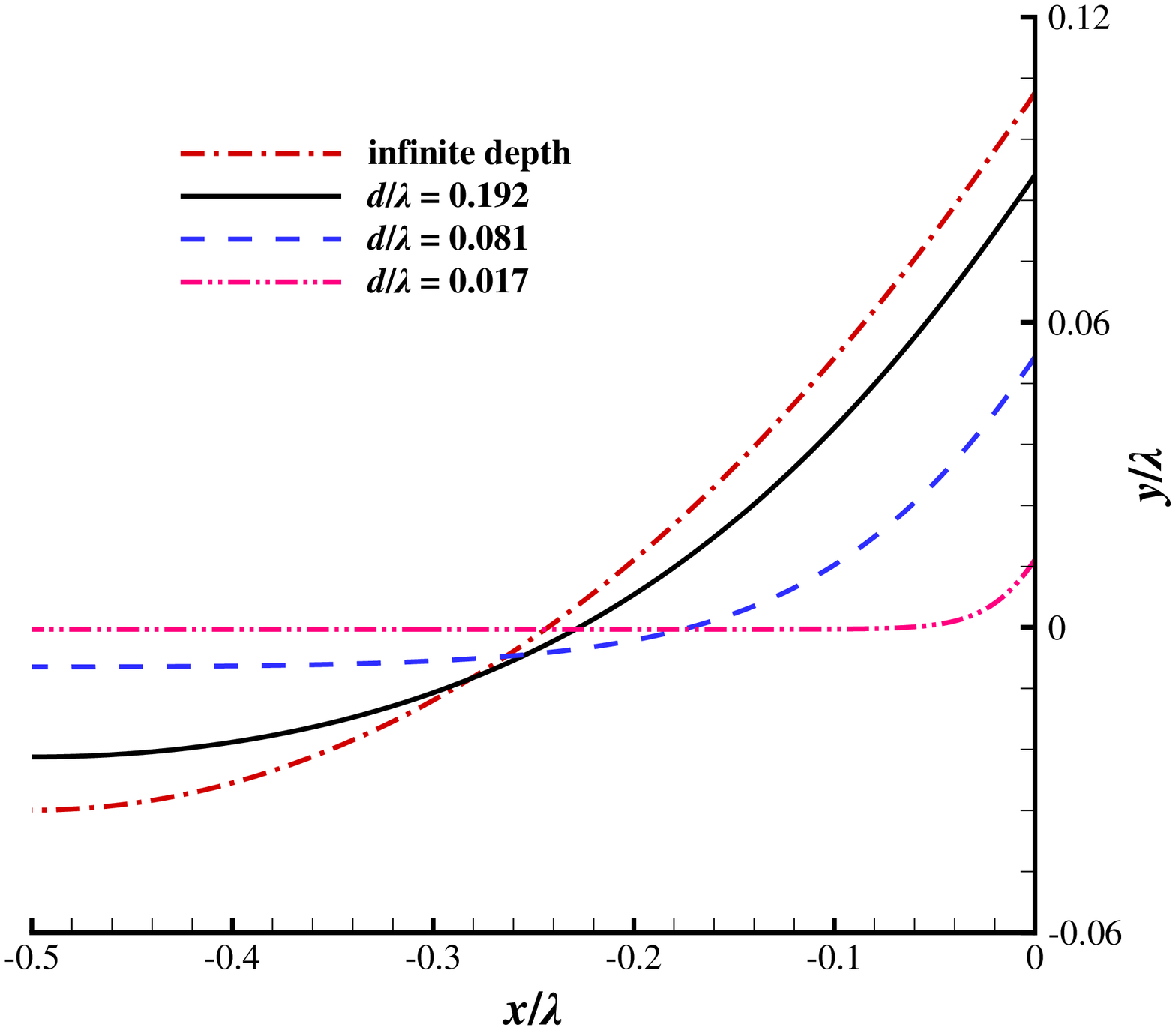}}
  \caption{Wave profiles in a variety of water depths, given by the first-order HAM-based iteration approach.}
\label{finite:profile:different:depth}
\end{figure}

\citet{Stokes1880Supplement}  gave a famous conjecture that the limiting wave (with extreme height) should have a sharp crest with an included angle $120^\circ$.   About one hundred years later, this conjecture was independently proved in mathematics by \citet{Amick1982On} and \citet{Plotnikov2002A} for Stokes' waves in arbitrary depth of water.    However, to the best of our knowledge, the detailed wave profiles for limiting Stokes' wave in extremely shallow water have not been reported.   Table~\ref{table:depth:angle} presents the included crest angles of the limiting Stokes' wave in a variety of water depths,  given by our iteration HAM approach.   All of the  included crest angles in different depth given by the HAM are very close to the theoretic value $120^\circ$.     The wave profiles in a variety of water depths given by the HAM  are  shown in Figure~\ref{finite:profile:different:depth}.
Note that the high-order Fourier coefficients $a_j$ play an important role in correctly describing the wave profile, especially the wave crest.  For instance, although Cokelet's perturbation method \citep{Cokelet1977Steep} can give $H/d$ with acceptable accuracy for $r_0<0.9$, however, it fails to give accurate wave profile even for $r_0>0.5$.   Fortunately, the HAM can always yield convergent results of {\em all} Fourier coefficients by means of choosing a proper convergence-control parameter $\hbar$.   This once again illustrates the superiority of the HAM over other methods  \citep{Schwartz1974Computer, Cokelet1977Steep}.

 \begin{table}
 \tabcolsep 0pt
\begin{center}
\def\temptablewidth{0.9\textwidth}
\begin{tabular*}{\temptablewidth}{@{\extracolsep{\fill}}ccc}
$r$  &  $H/\lambda$    & $(g\lambda)/(2\pi c^2)$ \\
21000 &   $1.3229\times 10^{-3}$& 60.410           \\
28000 &    $1.3251\times 10^{-3}$& 60.312         \\
35000 &    $1.3264\times 10^{-3}$& 60.249          \\
40000&   $1.3272\times 10^{-3}$& 60.214         \\
50000&  $1.3281\times 10^{-3}$& $60.175$          \\
55000&  $1.3281\times 10^{-3}$& $60.175$         \\
\end{tabular*}
\caption{Wave steepness $H/\lambda$ and wave speed parameter $(g\lambda)/\left(2\pi c^2\right)$ versus truncated terms $r$ in the case of $r_0=0.99$, given by the first-order HAM-based iteration approach.}   \label{table:different99:r}
 \end{center}
 \end{table}

 \begin{table}
 \tabcolsep 0pt
\begin{center}
\def\temptablewidth{0.5\textwidth}
\begin{tabular*}{\temptablewidth}{@{\extracolsep{\fill}}cc}
$r_0$  &  included crest angle    \\
0 &   $119.3^\circ$          \\
0.3 &    $119.2^\circ$       \\
0.6 &    $119.4^\circ$          \\
0.9&   $120.2^\circ$       \\
0.99&  $119.2^\circ$        \\
\end{tabular*}
\caption{Included crest angles in a variety of depths, given by the first-order HAM-based iteration approach.}   \label{table:depth:angle}
 \end{center}
 \end{table}

\begin{figure}
\centering
\subfigure[]{
\begin{minipage}[b]{5.6in}
\centering
\includegraphics[width=3.5in]{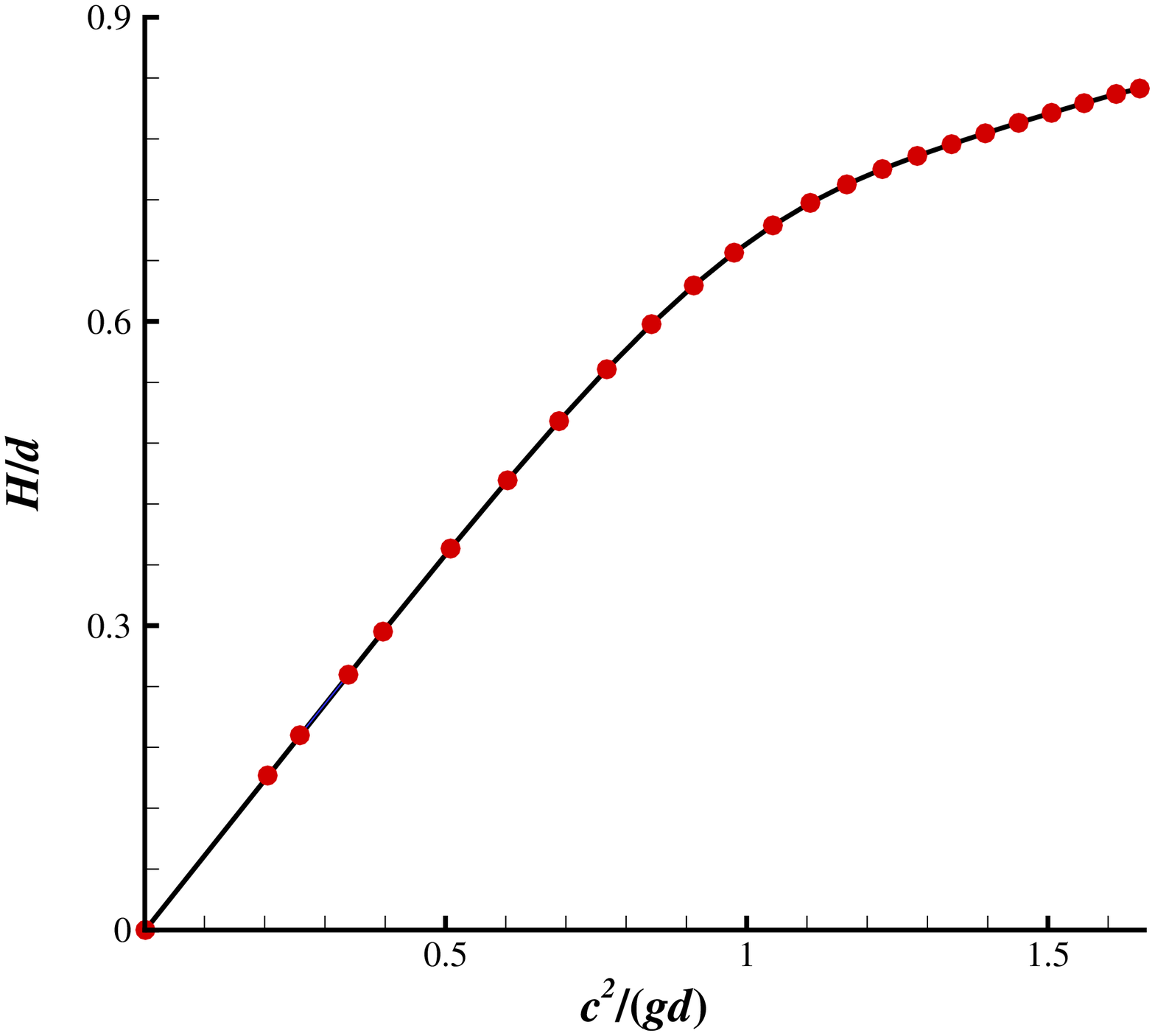}
\end{minipage}
}\\

\subfigure[]{
\begin{minipage}[b]{5.6in}
\centering
\includegraphics[width=3.5in]{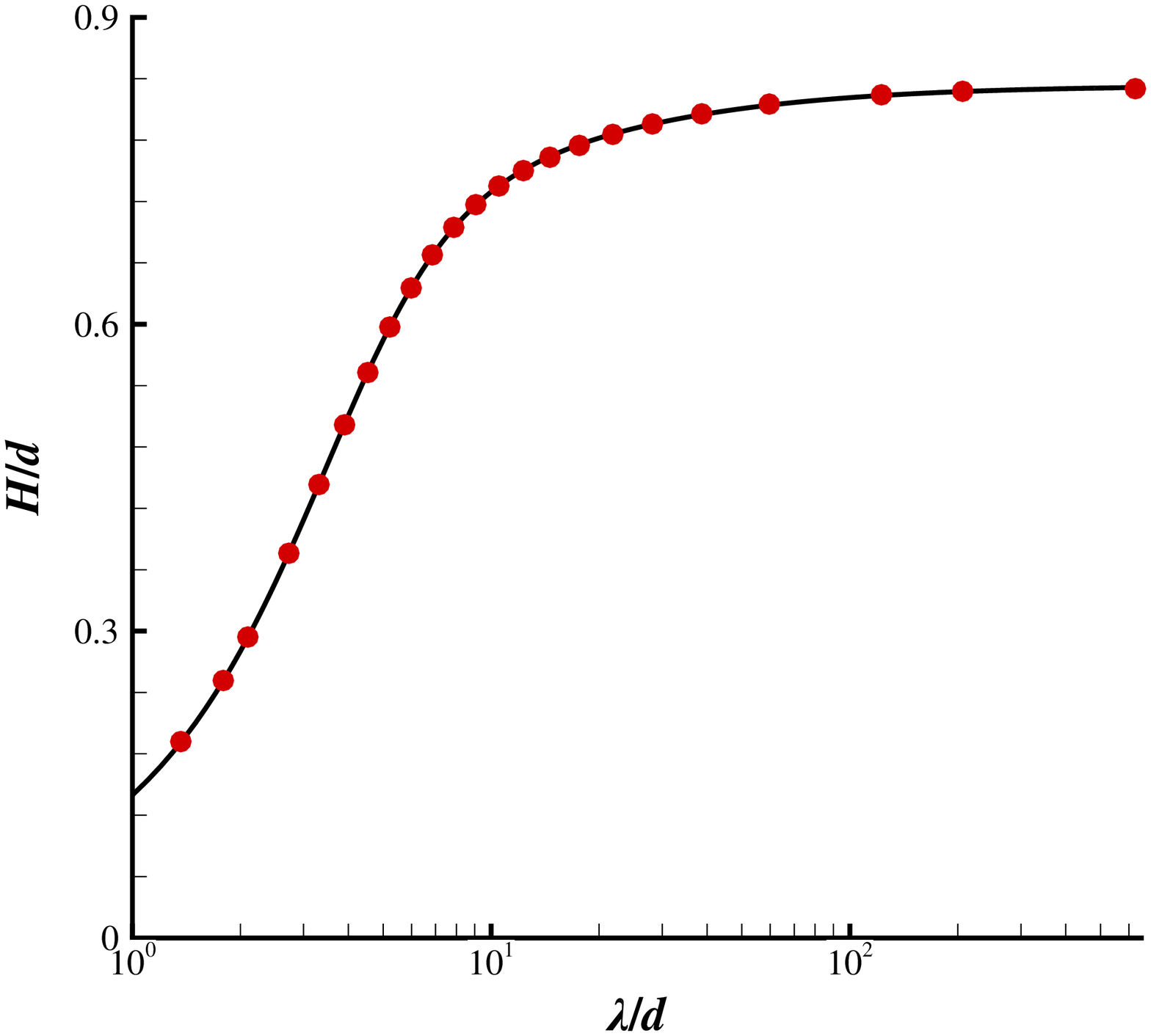}
\end{minipage}
}\caption{Comparison of $H/d$ given by the HAM-based iteration approach and fitted formulas (\ref{fit:HAM:1}), (\ref{fit:HAM:2}). $\bullet$, the first-order HAM-based iteration approach; (a) ---$\!$---, (\ref{fit:HAM:1});  (b) ---$\!$---, (\ref{fit:HAM:2}).} \label{figure:fit:HAM}
\end{figure}

According to the convergent results given by the iteration HAM approach,  we have the fitted formulas of $H/d$ versus $c^2/(gd)$ and $\lambda/d$:
\begin{eqnarray}
\frac{H}{d}&=&0.58557\;\frac{c^2}{gd}+0.62667\left(\frac{c^2}{gd}\right)^2-0.73410\left(\frac{c^2}{gd}\right)^3+0.19634\left(\frac{c^2}{gd}\right)^4,\label{fit:HAM:1}\\
\frac{H}{d}&=&\frac{0.14109\;\frac{\lambda}{d}+0.00804\left(\frac{\lambda}{d} \right)^2+0.00949\left(\frac{\lambda}{d} \right)^3}{1+0.09671\;\frac{\lambda}{d}+0.02695\left(\frac{\lambda}{d} \right)^2+0.01139\left(\frac{\lambda}{d} \right)^3},  \label{fit:HAM:2}
\end{eqnarray}
which agree quite well with our HAM results, as shown in Figure~\ref{figure:fit:HAM}.

Let us make some comparisons of the limiting Stokes'  waves given by different analytic/numerical methods.  Table \ref{table:compare:wavesteepness} presents the comparison of limiting wave steepness in a variety of depths.   The results of \citet{Schwartz1974Computer} are accurate only for $r_0<0.7$;  the results of \citet{Cokelet1977Steep} are accurate only for $r_0\leq0.8$;   the results of \citet{Williams1981Limiting} are of high accuracy for $r_0\leq 0.9$, although they are slightly smaller.   However, all of these methods fail to give convergent result for $r_0 > 0.9$, i.e., extremely shallow water.   Fortunately, the HAM can give convergent results for the limiting waves almost in {\em arbitrary} depth.

 \begin{table}
 \tabcolsep 0pt
\begin{center}
\def\temptablewidth{1\textwidth}
\begin{tabular*}{\temptablewidth}{@{\extracolsep{\fill}}ccccc}
$r_0$  &  \citet{Schwartz1974Computer}    & \citet{Cokelet1977Steep}  &   \citet{Williams1981Limiting} & the HAM\\
0 &   $1.4118\times 10^{-1}$ & $1.41055\times 10^{-1}$ & $1.41063\times 10^{-1}$   & $1.4108\times 10^{-1}$ \\
0.1 &   $1.380\times 10^{-1}$ &$1.378\times 10^{-1}$ & $1.37801\times 10^{-1}$  & $1.3782\times 10^{-1}$   \\
0.2 &   $1.285\times 10^{-1}$ & $1.285\times 10^{-1}$ & $1.28495\times 10^{-1}$   & $1.2851\times 10^{-1}$ \\
0.3 &  $1.145 \times 10^{-1}$ & $1.1443 \times 10^{-1}$ & $1.14439 \times 10^{-1}$ & $1.1446 \times 10^{-1}$ \\
0.4 &$9.75\times 10^{-2}$ & $9.739\times 10^{-2}$  & $9.7374\times 10^{-2}$&$9.7388\times 10^{-2}$  \\
0.5 &   $7.91\times 10^{-2}$ & $7.910\times 10^{-2}$ & $7.9072\times 10^{-2}$  & $7.9084\times 10^{-2}$ \\
0.6 & $6.14 \times 10^{-2}$ &$6.090 \times 10^{-2}$ &$6.0984 \times 10^{-2}$ & $6.0995 \times10^{-2}$ \\
0.7 &$4.5\times10^{-2}$  & $4.374\times10^{-2}$  &$4.3975\times10^{-2}$ & $4.3983\times10^{-2}$ \\
0.8 &-----&$2.79 \times 10^{-2}$&$2.8258 \times 10^{-2}$& $2.8263 \times 10^{-2}$       \\
0.9 &-----&$1.5 \times 10^{-2}$ &$1.3667 \times 10^{-2}$& $1.3670 \times 10^{-2}$       \\
0.95 &-----&----- &-----& $6.7292 \times 10^{-3}$       \\
0.97 &-----&----- &-----& $4.0128 \times 10^{-3}$       \\
0.99 &-----&----- &-----& $1.3281\times 10^{-3}$       \\
\end{tabular*}
\caption{Limiting wave steepness, $H/\lambda$, in a variety of depths.}   \label{table:compare:wavesteepness}
 \end{center}
 \end{table}

\begin{figure}
  \centerline{\includegraphics[width=3.5in]{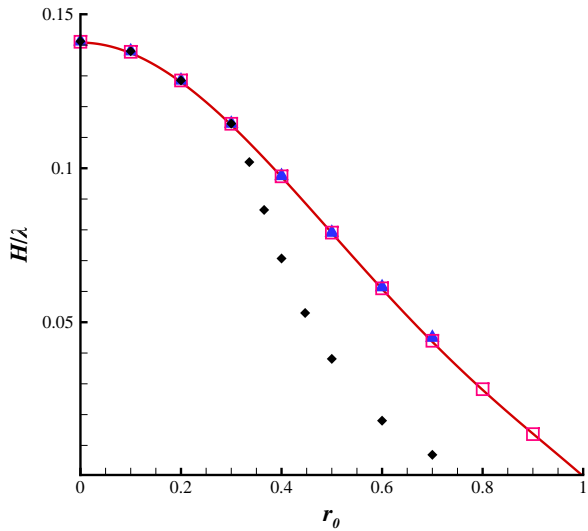}}
  \caption{Comparison of the limiting wave steepness $H/\lambda$. ---$\!$---, the first-order HAM-based iteration approach; $\blacklozenge$, perturbation method with the aid of Pad\'{e} approximants \citep{Schwartz1974Computer}; $\blacktriangle$, perturbation method with the aid of both Pad\'{e} approximants and Shanks's iterated $e_1$ transformation \citep{Schwartz1974Computer}; $\square$: Williams' numerical method \citep{Williams1981Limiting}.}
\label{Comparison:finite:height:depth}
\end{figure}

Figure \ref{Comparison:finite:height:depth} shows the comparison of the limiting wave steepness $H/\lambda$, given by \citet{Schwartz1974Computer}, \citet{Williams1981Limiting} and the HAM approach mentioned in this paper, respectively.  It is found that the perturbation method \citep{Schwartz1974Computer} is only valid for $r_0\in[0,0.7]$ even with the aid of extrapolation and Pad\'{e} approximant techniques;  Williams' numerical method \citep{Williams1981Limiting} is only valid for $r_0\in[0,0.9]$.  However, the HAM can give accurate convergent results even for $r_0\in [0,0.99]$.

\begin{figure}
  \centerline{\includegraphics[width=3.5in]{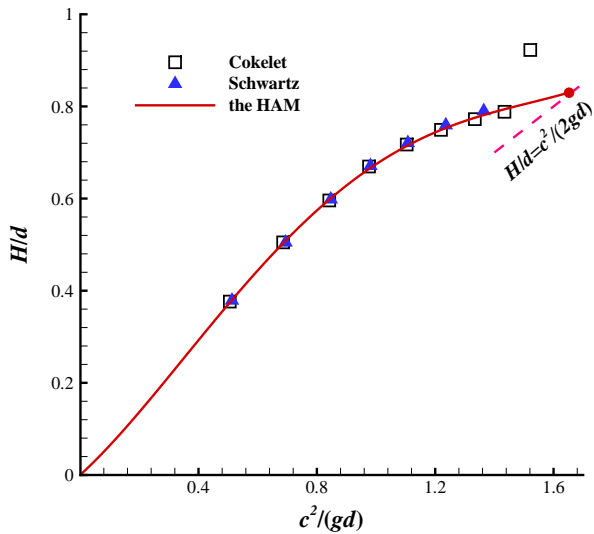}}
  \caption{Comparison of $H/d$, versus the squared Froude number, $c^2/(gd)$.  ---$\!$---, the first-order HAM-based iteration approach; $\bullet$, the case of $r_0=0.99$ given by the first-order HAM-based iteration approach; $\blacktriangle$, \citet{Schwartz1974Computer}; $\square$, \citet{Cokelet1977Steep}; --- --- ---, $H/d=c^2/(2gd)$.}
\label{limiting:dispersion:relation}
\end{figure}

Figure \ref{limiting:dispersion:relation} shows the comparison of $H/d$ versus the squared Froude number, $c^2/(gd)$.  It is found that Cokelet's perturbation method \citep{Cokelet1977Steep} fails in extremely shallow water, i.e., $r_0\geq0.9$.  By contrast, our results given by the HAM are valid almost in arbitrary water depth.  Besides, in the case of $r_0=0.99$, $H/d$ given by the HAM is in accord with the results of the highest solitary wave:
\begin{equation}
\left(\frac{H}{d} \right)_{max}=\frac{c^2}{2gd}   \quad \quad \mbox{ for  $r_0=1$}.
\end{equation}
This suggests that the solitary wave theory could be unified into the Stokes' wave theory.

\begin{figure}
  \centerline{\includegraphics[width=3.5in]{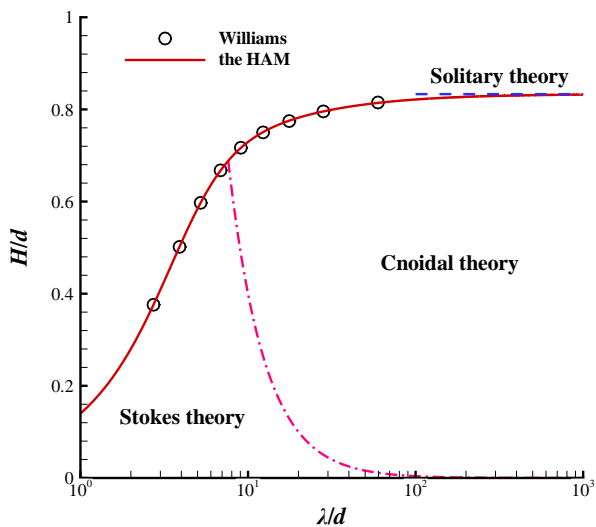}}
  \caption{Comparison of $H/d$, versus $\lambda/d$.  $\circ$, \citet{Williams1981Limiting}; ---$\!$---, the first-order HAM-based iteration approach; -- $\cdot$ --, demarcation line between Stokes and cnoidal theories, the Ursell number $H\lambda^2/d^3=40$ \citep{Hedges1995Regions}; -- -- --, $(H/d)_{max}=0.83322$ for solitary wave \citep{Hunter1983Accurate}.}
\label{Comparison:different:theory}
\end{figure}

According to \citet{Hedges1995Regions},  waves with the Ursell number $H\lambda^2/d^3>4000$ are regarded as solitary waves.  It is found that, in the case of $r_0=0.99$, corresponding to $\lambda/d\approx600$,  the $H\lambda^2/d^3$  of the limiting Stokes' wave given by the HAM reaches  $3\times10^5$.  Thus, the Stokes' wave theory is actually valid almost in \emph{arbitrary} depth, as shown in Figure~\ref{Comparison:different:theory}.   So,  in the frame of the HAM,  the Stokes wave theory can describe {\em not only} the periodic waves in deep and intermediate water {\em but also}  cnoidal wave in shallow water and solitary wave in extremely shallow water.

 In addition, the ratio of wave height to depth, $H/d$, of the highest solitary wave was widely studied by many researchers: $H/d=0.827$ was given by \citet{Yamada1957}, \citet{Lenau1966The}, \citet{Yamada1968On}, \citet{Longuethiggins1974On}; but $H/d=0.8332$ was given by \citet{Witting1981High}, \citet{Bergin1981High}, \citet{Williams1981Limiting}, \citet{ Hunter1983Accurate}.  Note that, $H/d=0.8303>0.827$ is given by the HAM in the case of $r_0=0.99$.  Hence the value $H/d=0.827$ for the highest solitary wave is denied by the HAM.

\begin{figure}
  \centerline{\includegraphics[width=3.5in]{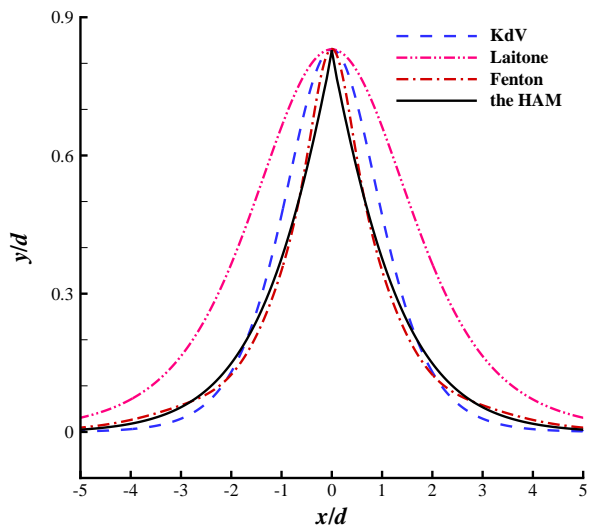}}
  \caption{Wave profile in the case of $r_0=0.99$.  -- -- --, exact solution of KdV equation \citep{KdV}; -- $\cdot\cdot$ --, Laitone's second order approximation solution \citep{Laitone1960The}; -- $\cdot$ --, Fenton's ninth-order approximation solution \citep{Fenton1972A}; ---$\!$---, homotopy approximation solution of equation (\ref{free:end}).}
\label{comparison:kdv:ham}
\end{figure}

Figure~\ref{comparison:kdv:ham} shows the wave profiles of the limiting wave in the case of $r_0=0.99$ given by the HAM from the exact wave equations, the KdV solution \citep{KdV},  Laitone's second-order solution \citep{Laitone1960The} and  Fenton's ninth-order solution \citep{Fenton1972A}.   It is found that only the HAM gives a wave profile with a sharply pointed crest, enclosing an angle $119.2^\circ$.  So, the KdV solution \citep{KdV}, Laitone's solution \citep{Laitone1960The} and Fenton's solution \citep{Fenton1972A} are all no longer valid in the limiting case.   However,  compared to the famous solitary solution of KdV equation \citep{KdV} and Laitone's solution \citep{Laitone1960The}, Fenton's ninth-order solution \citep{Fenton1972A} is  of higher accuracy.

In summary, using the iteration HAM approach with a proper convergence-control parameter, we gain limiting Stokes' waves almost in {\em arbitrary} water depth, {\em without} using any extrapolation techniques.  Therefore, in the frame of the HAM,  the Stokes' wave theory is a unified theory for {\em all} kinds of progressive waves in {\em arbitrary} depth, even including solitary waves in extremely shallow water.

\section{Concluding remarks}

Obviously, limiting Stokes' wave in shallow water  is a strong nonlinear problem.  Previous methods, especially the perturbation methods, usually suffer divergence either when the wave height approaches the peak value or when the water depth is extremely small.   For the limiting Stokes' wave, due to the existence of singularity locating exactly at the crest,  perturbation methods usually can gain convergent results  only for a small part of Fourier coefficients so that the extrapolation methods such as  Pad\'{e} approximant techniques and Shanks' transformation had to be used.

In this paper, we employ the homotopy analysis method (HAM) to solve the limiting Stokes' wave in arbitrary depth of water.  It is found that the convergence of {\em all} Fourier coefficients of the solutions can be guaranteed by choosing a proper  convergence-control parameter $\hbar$ in the frame of the HAM, as shown in Figure \ref{figure:c0:a1}.   In addition, since  the  Fourier series is used to represent the free surface with a sharp pointed crest,  using a large number of Fourier coefficients is inevitable.   For other analytic/numerical methods, this  might  lead to rather slow convergence of the Fourier coefficients of the solutions.   Fortunately,  the HAM also provides us great freedom to choose initial guesses of solutions.   Based on this kind of freedom of the HAM, we proposed an iteration HAM approach to greatly accelerate the convergence of {\em all} Fourier coefficients.   Note that, since we consider a large enough number of Fourier coefficients, and more importantly, all of these coefficients are convergent without using any extrapolation methods, hence we can obtain the accurate wave profile even in rather shallow water.

It should be emphasized that accurate representation of the wave profile in very shallow water is impossible for other methods, especially {\em without}   using any kind of extrapolation techniques.   For instance, although Cokelet's perturbation method \citep{Cokelet1977Steep} can give results of $H/\lambda$ with acceptable accuracy for $r_0<0.9$, however, it can only give a good wave profile for $r_0\leq0.5$.   Fortunately, by means of the HAM, we gain accurate limiting wave profiles in almost {\em arbitrary} depth of water, i.e., from $r_0 =0$ to $r_0=0.99$, {\em without} using any  extrapolation methods such as  Pad\'{e} approximant techniques and Shanks' transformation.   To the best of our  knowledge, accurate wave profile in the case of $r_0=0.99$ has been never reported.   This once again illustrates  the superiority of the HAM over  perturbation and traditional numerical methods for this famous problem.

According to \citet{Hedges1995Regions},  waves with the Ursell number $H\lambda^2/d^3>4000$ are regarded as solitary waves.  It is found that, in the case of $r_0=0.99$, corresponding to $\lambda/d\approx600$,  the $H\lambda^2/d^3$  of the Stokes' wave given by our HAM  approach  reaches  $3\times10^5$.  Thus, the Stokes' wave theory is actually valid almost in \emph{arbitrary} depth, as shown in Figure~\ref{Comparison:different:theory}.   So in the frame of the HAM,  the Stokes wave theory can describe {\em not only} the periodic waves in deep and intermediate water {\em but also}  cnoidal wave in shallow water and solitary wave in extremely shallow water.   Therefore, in the frame of the HAM, the Stokes' wave  is  a  unified  theory  for {\em all} kind of progressive waves, even including the limiting (extreme) solitary waves with a sharp crest of $120^\circ$  included angle in extremely shallow water!

Note that  the cubic relations between $a_j$ in equations (\ref{origin:equation:B})-(\ref{origin:equation:A}) were  considered in this paper, although the quadratic relations between the Fourier coefficients $a_j$ were reported by \citet{Higgins1978Some}.  Certainly, the computational efficiency could be improved by means of using the quadratic relations  \citep{Higgins1985Bifurcation, Balk1996A},   but one should obtain the same results as mentioned above in this paper, from a physical viewpoint.

From viewpoint of applied mathematics, this paper provides us an additional example to illustrate that the HAM can be indeed applied to find something completely new, such as the discovery of  the steady-state exactly/nearly resonant gravity waves with time-independent wave spectrum \citep{LIAO20111274,xu2012JFM,Liu2014JFM,Liu2015JFM,Liao2016JFM,Liu2018Finite}, or to attack  some challenging problems with high nonlinearity.

\section*{Acknowledgement}  Thanks to the anonymous reviewers for their valuable comments.   Thanks to Professor Yaosong Chen (Peking University, China) for his suggesting us to attack the limiting Stokes' wave in the extremely shallow water by means of the HAM.   This work was carried out on TH-2 at National Supercomputer Centre in Guangzhou, China.  It is partly supported by National Natural Science Foundation of China (Approval No. 11432009).

\appendix
\section{Detailed derivation of formulas (\ref{origin:equation:B})-(\ref{def:j:h})}
Rewrite (\ref{f:expression})
\begin{equation}
f(\zeta)=\sum_{i=-r}^r g_i \zeta^i,
\end{equation}
in which
\begin{eqnarray}
\left\{
\begin{split}
&g_i=a_{-i}r_0^{-2i}\quad\quad \mbox{when $i<0$ },\\
&g_0=a_0 ,\\
&g_i=a_{i}\quad\quad \mbox{when $i>0$}.     \label{def:gi}
\end{split}
\right.
\end{eqnarray}
Note that $R=1$ on the free surface, i.e., $\zeta=\mathrm{e}^{\mathrm{i}\theta}$.  We have
\begin{eqnarray}
f\;\bar{f}&=&\left(\sum_{i=-r}^r g_i \zeta^i \right)\left(\sum_{i=-r}^r g_i \zeta^{-i} \right)\nonumber\\
&=&\sum_{i=-r}^r g_i ^2 + \sum_{k=1}^{2r}\left[  \bigg{(}\zeta^k+\zeta^{-k}  \bigg{)}\left(\sum_{m=k-r}^{r}g_m g_{m-k}  \right) \right]\nonumber\\
&=&\sum_{i=-r}^r g_i ^2 + \sum_{k=1}^{2r}\left[  2\left(\sum_{m=k-r}^{r}g_m g_{m-k}  \right)\cos(k\theta) \right]\nonumber\\
&=&\sum_{k=0}^{2r}j_k\cos(k\theta),
\end{eqnarray}
where
\begin{equation}
j_0=\sum_{i=-r}^r g_i ^2, \quad\quad j_k=2\sum_{m=k-r}^r g_m g_{m-k},\quad\quad k=1,2,\cdots,2r.
\end{equation}
In addition, we have
\begin{eqnarray}
\int_0^{\theta}\mathrm{Im}[f]\mathrm{d}\theta&=&\int_0^{\theta}\left[\sum_{k=1}^r a_k \left(1-r_0^{2k} \right)\sin(k\theta)\right]\mathrm{d}\theta\nonumber\\
&=&\sum_{k=1}^r \frac{a_k \left(1-r_0^{2k} \right)}{k}-\sum_{k=1}^r \left[\frac{a_k \left(1-r_0^{2k}  \right)}{k} \cos(k\theta)\right] \nonumber\\
&=&\sum_{k=0}^{r}h_k\cos(k\theta),
\end{eqnarray}
where
\begin{equation}
h_0=\sum_{n_1=1}^r \frac{a_{n_1} \left(1-r_0^{2n_1} \right)}{n_1},\quad\quad h_n=- \frac{a_{n} \left(1-r_0^{2n} \right)}{n},\quad n=1,2,\cdots,r.
\end{equation}
Then we have
\begin{eqnarray}
&\;&f\;\bar{f}\;\int_0^{\theta}\mathrm{Im}[f]\mathrm{d}\theta\nonumber\\
&=&\left[\sum_{k=0}^{2r}j_k\cos(k\theta) \right]\left[\sum_{k=0}^{r}h_k\cos(k\theta) \right]\nonumber\\
&=&\left(j_0 h_0+\frac{1}{2}\sum_{n=1}^{r}j_n h_n\right)+{\cal N}_1\;\cos\theta +{\cal N}_2 \; \cos(2\theta)+\cdots,
\end{eqnarray}
in which ${\cal N}_1$, ${\cal N}_2$, $\cdots$, ${\cal N}_k$ are defined by (\ref{origin:equation:A}).

\bibliographystyle{jfm}
\bibliography{stokes}

\begin{thebibliography}{73}
\expandafter\ifx\csname natexlab\endcsname\relax\def\natexlab#1{#1}\fi
\def\au#1{#1} \def\ed#1{#1} \def\yr#1{#1}\def\at#1{#1}\def\jt#1{\textit{#1}}
  \def\bt#1{#1}\def\bvol#1{\textbf{#1}} \def\vol#1{#1} \def\pg#1{#1}
  \def\publ#1{#1}\def\arxiv#1{#1}\def\org#1{#1}\def\st#1{\textit{#1}}

\bibitem[Amick {\em et~al.\/}(1982)Amick, Fraenkel \& Toland]{Amick1982On}
{\sc \au{Amick, C.J.}, \au{Fraenkel, L.E.} \& \au{Toland, J.F.}} \yr{1982}
  \at{On the {S}tokes conjecture for the wave of extreme form}.  \jt{Acta
  Mathematica}  \bvol{148}~(1),  \pg{193--214}.

\bibitem[Balk(1996)]{Balk1996A}
{\sc \au{Balk, A.M.}} \yr{1996}  \at{A lagrangian for water waves}.
  \jt{Physics of Fluids}  \bvol{8}~(2),  \pg{416--420},  \arxiv{arXiv:
  https://doi.org/10.1063/1.868795}.

\bibitem[Byatt-Smith \& Longuet-Higgins(1976)]{Byatt1976On}
{\sc \au{Byatt-Smith, J.G.B.} \& \au{Longuet-Higgins, M.S.}} \yr{1976}  \at{On
  the {S}peed and {P}rofile of {S}teep {S}olitary {W}aves}.  \jt{Proceedings of
  the Royal Society A}  \bvol{350}~(1661),  \pg{175--189}.

\bibitem[Chandler \& Graham(1993)]{Chandler1993The}
{\sc \au{Chandler, G.~A.} \& \au{Graham, I.~G.}} \yr{1993}  \at{The
  {C}omputation of {W}ater {W}aves {M}odelled by {N}ekrasov’s {E}quation}.
  \jt{SIAM Journal on Numerical Analysis}  \bvol{30}~(4),  \pg{1041--1065}.

\bibitem[Chappelear(1961)]{Chappelear1961Direct}
{\sc \au{Chappelear, J.E.}} \yr{1961}  \at{Direct numerical calculation of wave
  properties}.  \jt{Journal of Geophysical Research}  \bvol{66}~(2),
  \pg{501--508}.

\bibitem[Chen \& Saffman(1980)]{Chen1980Numerical}
{\sc \au{Chen, B.} \& \au{Saffman, P.G.}} \yr{1980}  \at{Numerical evidence for
  the existence of new types of gravity waves of permanent form on deep water}.
   \jt{Studies in Applied Mathematics}  \bvol{62},  \pg{1--21}.

\bibitem[Cokelet(1977)]{Cokelet1977Steep}
{\sc \au{Cokelet, E.D.}} \yr{1977}  \at{{S}teep {G}ravity {W}aves in {W}ater of
  {A}rbitrary {U}niform {D}epth}.  \jt{Philosophical Transactions of the Royal
  Society of London}  \bvol{286}~(1335),  \pg{183--230}.

\bibitem[Crew \& Trinh(2016)]{Crew2016New}
{\sc \au{Crew, S.C.} \& \au{Trinh, P.H.}} \yr{2016}  \at{New singularities for
  {S}tokes waves}.  \jt{Journal of Fluid Mechanics}  \bvol{798},
  \pg{256--283}.

\bibitem[Dallaston \& Mccue(2010)]{Dallaston2010Accurate}
{\sc \au{Dallaston, M.C.} \& \au{Mccue, S.W.}} \yr{2010}  \at{Accurate series
  solutions for gravity-driven {S}tokes waves}.  \jt{Physics of Fluids}
  \bvol{22}~(8).

\bibitem[Davies(1951)]{davies1951the}
{\sc \au{Davies, T.V.}} \yr{1951}  \at{The {T}heory of {S}ymmetrical {G}ravity
  {W}aves of {F}inite {A}mplitude. {I}}.  \jt{Proceedings of The Royal Society
  A}  \bvol{208}~(1095),  \pg{475--486}.

\bibitem[Dean(1965)]{Dean1965Stream}
{\sc \au{Dean, R.G.}} \yr{1965}  \at{Stream {F}unction {R}epresentation of
  {N}onlinear {O}cean {W}aves}.  \jt{Journal of Geophysical Research
  Atmospheres}  \bvol{70}~(18),  \pg{4561--4572}.

\bibitem[Dyachenko {\em et~al.\/}(2014)Dyachenko, Lushnikov \&
  Korotkevich]{Dyachenko2014Complex}
{\sc \au{Dyachenko, S.A.}, \au{Lushnikov, P.M.} \& \au{Korotkevich, A.O.}}
  \yr{2014}  \at{Complex {S}ingularity of a {S}tokes {W}ave}.  \jt{Jetp
  Letters}  \bvol{98}~(11),  \pg{675--679}.

\bibitem[Dyachenko {\em et~al.\/}(2016)Dyachenko, Lushnikov \&
  Korotkevich]{Dyachenko2016Branch}
{\sc \au{Dyachenko, S.A.}, \au{Lushnikov, P.M.} \& \au{Korotkevich, A.O.}}
  \yr{2016}  \at{Branch {C}uts of {S}tokes {W}ave on {D}eep {W}ater. {P}art
  {I}: {N}umerical {S}olution and {P}ad{\'e} {A}pproximation}.  \jt{Studies in
  Applied Mathematics}  \bvol{137}~(4),  \pg{419--472}.

\bibitem[Fenton(1972)]{Fenton1972A}
{\sc \au{Fenton, J.D.}} \yr{1972}  \at{A ninth-order solution for the solitary
  wave}.  \jt{Journal of Fluid Mechanics}  \bvol{53}~(2),  \pg{257--271}.

\bibitem[Fenton(1988)]{Fenton1988The}
{\sc \au{Fenton, J.D.}} \yr{1988}  \at{The numerical solution of steady water
  wave problem}.  \jt{Computers \& Geosciences}  \bvol{14}~(3),  \pg{357--368}.

\bibitem[Fenton(1990)]{Fenton1990}
{\sc \au{Fenton, J.D.}} \yr{1990}  \at{Nonlinear {W}ave {T}heories}.  \jt{Ocean
  Engineering Science}  \bvol{9},  \pg{1--18}.

\bibitem[Grant(1973)]{Grant1973the}
{\sc \au{Grant, M.A.}} \yr{1973}  \at{The singularity at the crest of a finite
  amplitude progressive {S}tokes wave}.  \jt{Journal of Fluid Mechanics}
  \bvol{59}~(2),  \pg{257--262}.

\bibitem[Hedges(1995)]{Hedges1995Regions}
{\sc \au{Hedges, T.S.}} \yr{1995}  \at{Regions of validity of analytical wave
  theories}.  \jt{Ice Proceedings Water Maritime \& Energy}  \bvol{112}~(2),
  \pg{111--114}.

\bibitem[Hunter \& Vanden-Broeck(1983)]{Hunter1983Accurate}
{\sc \au{Hunter, J.K.} \& \au{Vanden-Broeck, J.M.}} \yr{1983}  \at{Accurate
  computations for steep solitary waves}.  \jt{Journal of Fluid Mechanics}
  \bvol{136}~(136),  \pg{63--71}.

\bibitem[Karabut(1998)]{Karabut1998An}
{\sc \au{Karabut, E.A.}} \yr{1998}  \at{An approximation for the highest
  gravity waves on water of finite depth}.  \jt{Journal of Fluid Mechanics}
  \bvol{372},  \pg{45--70}.

\bibitem[Kimiaeifar {\em et~al.\/}(2011)Kimiaeifar, Lund, Thomsen \&
  Sørensen]{Kimiaeifar2011Application}
{\sc \au{Kimiaeifar, A.}, \au{Lund, E.}, \au{Thomsen, O.T.} \& \au{Sørensen,
  J.D.}} \yr{2011}  \at{Application of the homotopy analysis method to
  determine the analytical limit state functions and reliability index for
  large deflection of a cantilever beam subjected to static co-planar loading}.
   \jt{Computers \& Mathematics with Applications}  \bvol{62}~(12),
  \pg{4646--4655}.

\bibitem[Klopman(1990)]{Klopman1990A}
{\sc \au{Klopman, G.}} \yr{1990}  \at{A note on integral properties of periodic
  gravity waves in the case of a non-zero mean {E}ulerian velocity}.
  \jt{Journal of Fluid Mechanics}  \bvol{211}~(5),  \pg{609--615}.

\bibitem[Korteweg \& de~Vries(1895)]{KdV}
{\sc \au{Korteweg, D.J.} \& \au{de~Vries, G.}} \yr{1895}  \at{On the change of
  form of long waves advancing in a rectangular canal and on a new type of long
  stationary waves}.  \jt{Phil. Mag.}  \bvol{39},  \pg{422--443}.

\bibitem[Laitone(1960)]{Laitone1960The}
{\sc \au{Laitone, E.V.}} \yr{1960}  \at{The second approximation to cnoidal and
  solitary waves}.  \jt{Journal of Fluid Mechanics}  \bvol{9}~(3),
  \pg{430--444}.

\bibitem[Lenau(1966)]{Lenau1966The}
{\sc \au{Lenau, C.W.}} \yr{1966}  \at{The solitary wave of maximum amplitude}.
  \jt{Journal of Fluid Mechanics}  \bvol{26}~(2),  \pg{309--320}.

\bibitem[Liao(1992)]{LiaoPhd}
{\sc \au{Liao, S.J.}} \yr{1992}  \at{Proposed homotopy analysis techniques for
  the solution of nonlinear problem}. {PhD thesis}, Shanghai Jiao Tong
  University.

\bibitem[Liao(1999)]{Liao1999A}
{\sc \au{Liao, S.J.}} \yr{1999}  \at{A {U}niformly {V}alid {A}nalytic
  {S}olution of 2{D} {V}iscous {F}low {P}ast a {S}emi-{I}nfinite {F}lat
  {P}late}.  \jt{Journal of Fluid Mechanics}  \bvol{385},  \pg{101--128}.

\bibitem[Liao(2003)]{Liaobook}
{\sc \au{Liao, S.J.}} \yr{2003} {\em {Beyond Perturbation: Introduction to
  Homotopy Analysis Method}\/}.  \publ{Boca Raton: Chapman \& Hall/CRC}.

\bibitem[Liao(2009)]{Liao2009}
{\sc \au{Liao, S.J.}} \yr{2009}  \at{Notes on the homotopy analysis method:
  Some definitions and theorems}.  \jt{Commun. Nonlinear Sci. Numer. Simul.}
  \bvol{14}~(4),  \pg{983--997}.

\bibitem[Liao(2010)]{Liao2010}
{\sc \au{Liao, S.J.}} \yr{2010}  \at{An optimal homotopy-analysis approach for
  strongly nonlinear differential equations}.  \jt{Commun. Nonlinear Sci.
  Numer. Simul.}  \bvol{15}~(8),  \pg{2003--2016}.

\bibitem[Liao(2011)]{LIAO20111274}
{\sc \au{Liao, S.J.}} \yr{2011}  \at{On the homotopy multiple-variable method
  and its applications in the interactions of nonlinear gravity waves}.
  \jt{Communications in Nonlinear Science and Numerical Simulation}
  \bvol{16}~(3),  \pg{1274 -- 1303}.

\bibitem[Liao(2012)]{liaobook2}
{\sc \au{Liao, S.J.}} \yr{2012} {\em Homotopy {A}nalysis {M}ethod in
  {N}onlinear {D}ifferential {E}quations\/}.  \publ{New York: Springer-Verlag}.

\bibitem[Liao {\em et~al.\/}(2016)Liao, Xu \& Stiassnie]{Liao2016JFM}
{\sc \au{Liao, S.J.}, \au{Xu, D.L.} \& \au{Stiassnie, M.}} \yr{2016}  \at{On
  the steady-state nearly resonant waves}.  \jt{J. Fluid Mech.}  \bvol{794},
  \pg{175--199}.

\bibitem[Liu \& Liao(2014)]{Liu2014JFM}
{\sc \au{Liu, Z.} \& \au{Liao, S.J.}} \yr{2014}  \at{Steady-state resonance of
  multiple wave interactions in deep water}.  \jt{J. Fluid Mech.}  \bvol{742},
  \pg{664--700}.

\bibitem[Liu {\em et~al.\/}(2015)Liu, Xu, Li, Peng, Alsaedi \&
  Liao]{Liu2015JFM}
{\sc \au{Liu, Z.}, \au{Xu, D.L.}, \au{Li, J.}, \au{Peng, T.}, \au{Alsaedi, A.}
  \& \au{Liao, S.J.}} \yr{2015}  \at{On the existence of steady-state resonant
  waves in experiments}.  \jt{J. Fluid Mech.}  \bvol{763},  \pg{1--23}.

\bibitem[Liu {\em et~al.\/}(2018{\natexlab{{\em a\/}}})Liu, Xu \&
  Liao]{Liu2018Finite}
{\sc \au{Liu, Z.}, \au{Xu, D.L.} \& \au{Liao, S.J.}} \yr{2018{\natexlab{{\em
  a\/}}}}  \at{Finite amplitude steady-state wave groups with multiple near
  resonances in deep water}.  \jt{Journal of Fluid Mechanics}  \bvol{835},
  \pg{624--653}.

\bibitem[Liu {\em et~al.\/}(2018{\natexlab{{\em b\/}}})Liu, Xu \&
  Liao]{Liu2018Mass}
{\sc \au{Liu, Z.}, \au{Xu, D.L.} \& \au{Liao, S.J.}} \yr{2018{\natexlab{{\em
  b\/}}}}  \at{Mass, momentum and energy flux conservation between linear and
  nonlinear steady-state wave groups}.  \jt{Physics of Fluids} Accepted.

\bibitem[Longuet-Higgins(1975)]{Longuet1975Integral}
{\sc \au{Longuet-Higgins, M.S.}} \yr{1975}  \at{Integral {P}roperties of
  {P}eriodic {G}ravity {W}aves of {F}inite {A}mplitude}.  \jt{Proceedings of
  the Royal Society of London}  \bvol{342}~(1629),  \pg{157--174}.

\bibitem[Longuet-Higgins(1978)]{Higgins1978Some}
{\sc \au{Longuet-Higgins, M.S.}} \yr{1978}  \at{Some {N}ew {R}elations
  {B}etween {S}tokes's {C}oefficients in the {T}heory of {G}ravity {W}aves}.
  \jt{IMA Journal of Applied Mathematics}  \bvol{22}~(3),  \pg{261--273}.

\bibitem[Longuet-Higgins(1985)]{Higgins1985Bifurcation}
{\sc \au{Longuet-Higgins, M.S.}} \yr{1985}  \at{Bifurcation in gravity waves}.
  \jt{Journal of Fluid Mechanics}  \bvol{151},  \pg{457--475}.

\bibitem[Longuet-Higgins \& Fenton(1974)]{Longuethiggins1974On}
{\sc \au{Longuet-Higgins, M.S.} \& \au{Fenton, J.D.}} \yr{1974}  \at{On the
  {M}ass, {M}omentum, {E}nergy and {C}irculation of a {S}olitary wave. {II}}.
  \jt{Proceedings of the Royal Society A Mathematical Physical \& Engineering
  Sciences}  \bvol{A340},  \pg{471--493}.

\bibitem[Longuet-Higgins \& Fox(1978)]{Longuet-higgins1978Theory}
{\sc \au{Longuet-Higgins, M.S.} \& \au{Fox, M.J.H.}} \yr{1978}  \at{Theory of
  the almost-highest wave. {P}art 2. {M}atching and analytic extension}.
  \jt{Journal of Fluid Mechanics}  \bvol{85}~(4),  \pg{769 -- 786}.

\bibitem[Lushnikov(2016)]{Lushnikov2016Structure}
{\sc \au{Lushnikov, P.M.}} \yr{2016}  \at{Structure and location of branch
  point singularities for {S}tokes waves on deep water}.  \jt{Journal of Fluid
  Mechanics}  \bvol{800},  \pg{557--594}.

\bibitem[Lushnikov {\em et~al.\/}(2017)Lushnikov, Dyachenko \&
  Silantyev]{Lushnikov2017New}
{\sc \au{Lushnikov, P.M.}, \au{Dyachenko, S.A.} \& \au{Silantyev, D.A.}}
  \yr{2017}  \at{New conformal mapping for adaptive resolving of the complex
  singularities of {S}tokes wave.}  \jt{Proceedings of the Royal Society A}
  \bvol{473}, ( DOI: 10.1098/rspa.2017.0198).

\bibitem[Mastroberardino(2011)]{Mastroberardino2011Homotopy}
{\sc \au{Mastroberardino, A.}} \yr{2011}  \at{Homotopy analysis method applied
  to electrohydrodynamic flow}.  \jt{Communications in Nonlinear Science \&
  Numerical Simulation}  \bvol{16}~(7),  \pg{2730--2736}.

\bibitem[Michell(1893)]{Michell1893}
{\sc \au{Michell, J.H.}} \yr{1893}  \at{The highest waves in water}.
  \jt{Philos. Mag.}  \bvol{36}~(5),  \pg{430--437}.

\bibitem[Nekrasov(1920)]{Nekrasov1920}
{\sc \au{Nekrasov, A.I.}} \yr{1920}  \at{On {S}tokes' wave}.  \jt{Isv.
  Ivanovo-Voznesesk. Politekhn}  \pg{pp. 81--89}.

\bibitem[Olfe \& Rottman(1980)]{Olfe1980Some}
{\sc \au{Olfe, D.B.} \& \au{Rottman, J.W.}} \yr{1980}  \at{Some new
  highest-wave solutions for deep-water waves of permanent form}.  \jt{Journal
  of Fluid Mechanics}  \bvol{100}~(4),  \pg{801--810}.

\bibitem[Plotnikov(2002)]{Plotnikov2002A}
{\sc \au{Plotnikov, P.I.}} \yr{2002}  \at{A {P}roof of the {S}tokes
  {C}onjecture in the {T}heory of {S}urface {W}aves *}.  \jt{Studies in Applied
  Mathematics}  \bvol{108}~(2),  \pg{217--244}, translated from Dinamika
  Sploshn. Sredy 57 (1982), 41-76.

\bibitem[Rienecker \& Fenton(1981)]{Rienecker1981A}
{\sc \au{Rienecker, M.M.} \& \au{Fenton, J.D.}} \yr{1981}  \at{A {F}ourier
  approximation method for steady water waves}.  \jt{Journal of Fluid
  Mechanics}  \bvol{104}~(104),  \pg{119--137}.

\bibitem[Sardany{\'e}s {\em et~al.\/}(2015)Sardany{\'e}s, Rodrigues,
  Janu{\'a}rio, Martins, Gil-G{\'o}mez \& Duarte]{Duarte2015}
{\sc \au{Sardany{\'e}s, J.}, \au{Rodrigues, C.}, \au{Janu{\'a}rio, C.},
  \au{Martins, N.}, \au{Gil-G{\'o}mez, G.} \& \au{Duarte, J.}} \yr{2015}
  \at{Activation of effector immune cells promotes tumor stochastic extinction:
  A homotopy analysis approach}.  \jt{Appl. Math. Comput.}  \bvol{252},
  \pg{484 -- 495}.

\bibitem[Schwartz(1972)]{Schwartz1972}
{\sc \au{Schwartz, L.W.}} \yr{1972}  \at{{A}nalytic continuation of {S}tokes'
  expansion for gravity waves}. {PhD thesis}, Stanford University.

\bibitem[Schwartz(1974)]{Schwartz1974Computer}
{\sc \au{Schwartz, L.W.}} \yr{1974}  \at{Computer extension and analytic
  continuation of {S}tokes' expansion for gravity waves}.  \jt{Journal of Fluid
  Mechanics}  \bvol{62}~(3),  \pg{553--578}.

\bibitem[Schwartz \& Fenton(1982)]{And1982Strongly}
{\sc \au{Schwartz, L.W.} \& \au{Fenton, J.D.}} \yr{1982}  \at{Strongly
  {N}onlinear {W}aves}.  \jt{Annual Review of Fluid Mechanics}  \bvol{14},
  \pg{39--60}.

\bibitem[Shanks(1954)]{Shanks1954Non}
{\sc \au{Shanks, D.}} \yr{1954}  \at{Non-linear {T}ransformations of
  {D}ivergent and {S}lowly {C}onvergent {S}equences}.  \jt{Journal of
  Mathematics \& Physics}  \bvol{34}~(1),  \pg{1--42}.

\bibitem[Stokes(1847)]{StokesOn}
{\sc \au{Stokes, G.G.}} \yr{1847}  \at{On the {T}heory of {O}scillatory
  {W}aves}.  \jt{Trans Cambridge Philos Soc}  \bvol{8},  \pg{441--455}.

\bibitem[Stokes(1880)]{Stokes1880Supplement}
{\sc \au{Stokes, G.G.}} \yr{1880}  \at{Supplement to a paper on the theory of
  oscillatory waves}.  \jt{Mathematical \& Physical Papers} ~(1),
  \pg{314--326}.

\bibitem[Sulem {\em et~al.\/}(1983)Sulem, Sulem \& Frisch]{Sulem1983Tracing}
{\sc \au{Sulem, Catherine}, \au{Sulem, Pierre~Louis} \& \au{Frisch, Hélène}}
  \yr{1983}  \at{Tracing complex singularities with spectral methods}.
  \jt{Journal of Computational Physics}  \bvol{50}~(1),  \pg{138--161}.

\bibitem[Tanveer(1991)]{Tanveer137}
{\sc \au{Tanveer, S.}} \yr{1991}  \at{Singularities in water waves and
  {R}ayleigh-{T}aylor instability}.  \jt{Proceedings of the Royal Society of
  London A: Mathematical, Physical and Engineering Sciences}
  \bvol{435}~(1893),  \pg{137--158}.

\bibitem[Tulin(1983)]{Tulin1983An}
{\sc \au{Tulin, M.P.}} \yr{1983}  \at{An exact theory of gravity wave
  generation by moving bodies, its approximation and its implication}.
  \jt{14th Symposium on Naval Hydrodynamics, Academic Press, New York}  \pg{pp.
  19--51}.

\bibitem[Vajravelu \& {Van}~Gorder(2012)]{KV2012}
{\sc \au{Vajravelu, K.} \& \au{{Van}~Gorder, R.~A.}} \yr{2012} {\em Nonlinear
  {F}low {P}henomena and {H}omotopy {A}nalysis: {F}luid {F}low and {H}eat
  {T}ransfer\/}.  \publ{Heidelberg: Springer}.

\bibitem[Van~Gorder \& Vajravelu(2008)]{KV2008}
{\sc \au{Van~Gorder, R.A.} \& \au{Vajravelu, K.}} \yr{2008}  \at{Analytic and
  numerical solutions to the {L}ane-{E}mden equation}.  \jt{Physics Letters A}
  \bvol{372},  \pg{6060--6065}.

\bibitem[Vanden-Broeck(1986)]{Vanden1986Steep}
{\sc \au{Vanden-Broeck, J.M.}} \yr{1986}  \at{Steep gravity waves: {H}avelock
  method revisited}  \bvol{29}~(9),  \pg{3084--3085}.

\bibitem[Vanden-Broeck \& Schwartz(1979)]{J1979Numerical}
{\sc \au{Vanden-Broeck, J.M.} \& \au{Schwartz, L.W.}} \yr{1979}  \at{Numerical
  computation of steep gravity waves in shallow water}.  \jt{Physics of Fluids}
   \bvol{22}~(10),  \pg{1868--1871}.

\bibitem[Vanden-Broeck \& Miloh(1995)]{Vanden1995Computations}
{\sc \au{Vanden-Broeck, J.~M.} \& \au{Miloh, T.}} \yr{1995}  \at{Computations
  of {S}teep {G}ravity {W}aves by a {R}efinement of {D}avies–{T}ulin’s
  {A}pproximation}.  \jt{Siam Journal on Applied Mathematics}  \bvol{55}~(4),
  \pg{892--903}.

\bibitem[Williams(1981)]{Williams1981Limiting}
{\sc \au{Williams, J.M.}} \yr{1981}  \at{Limiting {G}ravity {W}aves in {W}ater
  of {F}inite {D}epth}.  \jt{Philosophical Transactions of the Royal Society A
  Mathematical Physical \& Engineering Sciences}  \bvol{302}~(1466),
  \pg{139--188}.

\bibitem[Witting(1981)]{Witting1981High}
{\sc \au{Witting, J.M.}} \yr{1981}  \at{High solitary waves in water: {R}esults
  of calculations}.  \jt{NRL Rep.} .

\bibitem[Witting \& Bergin(1981)]{Bergin1981High}
{\sc \au{Witting, J.M.} \& \au{Bergin, J.M.}} \yr{1981}  \at{High {S}olitary
  {W}aves in {W}ater: {A} {R}efined {N}umerical {M}ethod}.  \jt{NRL Rep.} .

\bibitem[Xu {\em et~al.\/}(2012)Xu, Lin, Liao \& Stiassnie]{xu2012JFM}
{\sc \au{Xu, D.L.}, \au{Lin, Z.L.}, \au{Liao, S.J.} \& \au{Stiassnie, M.}}
  \yr{2012}  \at{On the steady-state fully resonant progressive waves in water
  of finite depth}.  \jt{J. Fluid Mech.}  \bvol{710},  \pg{379--418}.

\bibitem[Yamada(1957)]{Yamada1957}
{\sc \au{Yamada, H.}} \yr{1957}  \at{Highest waves of permanent type on the
  surface of deep water}.  \jt{Appl. Mech. Res. Rep., Kyushu Univ.}
  \bvol{5}~(18),  \pg{37--52}.

\bibitem[Yamada \& Shiotani(1968)]{Yamada1968On}
{\sc \au{Yamada, H.} \& \au{Shiotani, T.}} \yr{1968}  \at{On the {H}ighest
  {W}ater {W}aves of {P}ermanent {T}ype}.  \jt{Modern Healthcare}
  \bvol{34}~(38),  \pg{17}.

\bibitem[Zhong \& Liao(2017)]{Zhong2017}
{\sc \au{Zhong, X.X.} \& \au{Liao, S.J.}} \yr{2017}  \at{{Analytic {S}olutions
  of {Von K{\'a}rm{\'a}n} {P}late under {A}rbitrary {U}niform {P}ressure --
  {E}quations in {D}ifferential {F}orm}}.  \jt{Stud. Appl. Math.}  \bvol{138},
  \pg{371--400}, ( DOI: 10.1111/sapm.12158).

\bibitem[Zhong \& Liao(2018)]{Zhong2018Analytic}
{\sc \au{Zhong, X.X.} \& \au{Liao, S.J.}} \yr{2018}  \at{{Analytic {S}olutions
  of {Von K{\'a}rm{\'a}n} {P}late under {A}rbitrary {U}niform {P}ressure --
  {E}quations in {I}ntegral {F}orm}}.  \jt{Science China Physics, Mechanics \&
  Astronomy}  \bvol{61},  \pg{014611}.

\end{thebibliography}

\end{document}